\documentclass[fleqn,usenatbib]{mnras}
\usepackage{graphicx}	% Including figure files
\usepackage{amsmath}	% Advanced maths commands
\usepackage{amssymb}	% Extra maths symbols
\usepackage{xspace}

\newcommand{\mufasa}{\mbox{{\sc \small Mufasa}}\xspace}
\newcommand{\simba}{\mbox{{\sc \small Simba}}\xspace}
\newcommand{\MUV}{{\rm M_{1500}}}

\newcommand{\hmpc}{h^{-1}{\rm Mpc}}

\newcommand{\hst}{{\it HST}}
\newcommand{\jwst}{{\it JWST}}

\newcommand{\tmp}[1]{{\color{black}{#1}}}

\graphicspath{{./}{figures/}}

\title[Simba EoR Galaxies]{Photometric properties of reionization-epoch galaxies in the Simba simulations}

\author[X. H. Wu et al.]{
Xiaohan Wu$^{1}$\thanks{E-mail: xiaohan.wu@cfa.harvard.edu}\href{https://orcid.org/0000-0003-2061-4299}{\includegraphics[scale=0.8]{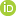}},
Romeel Dav{\'e}$^{2,3,4}$,
Sandro Tacchella$^{1}$\href{https://orcid.org/0000-0002-8224-4505}{\includegraphics[scale=0.8]{ORCIDiD_icon16x16.png}},
Jennifer Lotz$^{5}$
\\
$^{1}$Center for Astrophysics | Harvard \& Smithsonian, 60 Garden St, Cambridge, MA 02138, USA\\
$^{2}$Institute for Astronomy, University of Edinburgh, EH9 3HJ, UK\\
$^3$University of the Western Cape, Bellville, Cape Town 7535, South Africa\\
$^4$South African Astronomical Observatories, Observatory, Cape Town 7925, South Africa\\
$^{5}$Space Telescope Science Institute, 3700 San Martin Drive, Baltimore, MD 21218, USA
}

\date{Accepted XXX. Received YYY; in original form ZZZ}

\pubyear{2018}

\begin{document}
\label{firstpage}
\pagerange{\pageref{firstpage}--\pageref{lastpage}}
\maketitle

\begin{abstract}
We study the photometric properties and sizes of the reionization-epoch galaxies in high-resolution \simba\ cosmological hydrodynamical simulations with box sizes of $[25,50]\hmpc$. Assuming various attenuation laws, we compute photometry by extincting each star particle's spectrum using the line-of-sight gas metal column density. The predicted ultraviolet luminosity function (UVLF) generally agrees with observations at $z=6$, owing to a partial cancellation between the high metallicities of the simulated galaxies and lower dust-to-metal ratios. The simulated $z=8$ UVLF is low compared to observations, likely owing to excessive dust extinction. \simba\ predicts UV continuum slopes ($\beta$) in agreement with the $z=6$ observations, with the best agreement obtained using a Calzetti extinction law. Interestingly, the gas-phase mass-metallicity relation in \simba\ is higher at $z\sim 6$ than at $z\sim 2$, suggesting that rapid early enrichment (and dust growth) \tmp{might be} necessary to match the observed $\beta$. We find that $\beta$ is more sensitive to the dust extinction law than the UVLF. By generating mock {\it James Webb Space Telescope} (\jwst) images and analysing in a manner similar to observations, we show that \simba's galaxy size--luminosity relation well reproduces the current $z=6$ {\it Hubble} observations. Unlike observations at lower redshifts, \simba\ predicts similar rest-UV and rest-optical sizes of $z=6$ galaxies, owing to weak age gradients and dust extinction in star-forming regions counteract each other to weaken the color gradients within galaxies. These predictions will be testable with \jwst.
\end{abstract}


%%%%%%%%%%%%%%%%%%%%%%%%%%%%%%%%%%%%%%%%%%%%%%%%%%

%%%%%%%%%%%%%%%%% BODY OF PAPER %%%%%%%%%%%%%%%%%%

\section{Introduction}
\label{sec:intro}

% need more citations to observational papers?
Our understanding of galaxy evolution over the first billion years after the Big Bang has been revolutionized since the installation of the Wide Field Camera 3 (WFC3) on board the {\it Hubble Space Telescope} (\hst). The remarkable near-infrared (IR) capabilities of WFC3 has enabled more than 1000 galaxies to be detected at $6<z<10$, allowing a thorough investigation into star formation in the very early universe. This cosmic frontier will be pushed even more forward with the launch of the {\it James Webb Space Telescope} (\jwst), with its unprecedented IR sensitivity and spatial resolution. The physical properties of these young galaxies provide direct tests of our theoretical picture of galaxy formation, where gas inflows, outflows, and other feedback processes shape the star formation histories together. In addition, observations of these early objects shed light on whether their emergence provided enough ionizing photons to make reionization of the intergalactic medium (IGM) possible -- the last major phase transition of the IGM \citep[for a recent review, see][]{Stark16}.

One of the most fundamental measurements at $z\gtrsim6$ is the galaxy rest-frame UV luminosity function (UVLF). The UVLF gives the volume density of galaxies as a function of their rest-UV luminosity, canonically parametrized by a \citet{Sche76} function. Comparisons between the galaxy LF and the halo mass function provide insights into the physical processes that affect galaxy formation across cosmic time, e.g. supernova feedback which shapes the faint-end of the LF \citep[e.g.][]{Somerv15}. Moreover, observing the UVLF at $z\gtrsim6$ probes the role of early galaxies in the epoch of reionization (EoR), owing to the correlation between the rest-frame UV luminosity and recent star formation activities for all but the most dust-obscured galaxies. The recent measurements of \citet{Bouw15} and \citet{Fink15} showed that the faint-end slope of the UVLF is steeper with increasing redshift, consistent with the evolution of the halo mass function \citep[see also][]{Tacc13, Tacc18, Mason15}. It reaches $\simeq-2$ by $z\simeq7-8$ \citep[but see also][]{Bouw17, Live17, Atek18}, suggesting that low-luminosity galaxies dominate the integrated UV luminosity during reionization. Such a steep faint end slope also implies that the ionizing photon budget produced by these early galaxies is sufficient to reionize the universe by $z\simeq6$ with modest ionising photon escape fractions \citealp[(e.g.][]{Robe13, Fink19}\citealp[; but see also][]{Naidu19}).

While the UVLF gives the galaxy abundances, the UV continuum slope $\beta$ (defined by $f_\lambda \propto \lambda^\beta$) provides information about the stellar populations and the interstellar medium of the galaxies. Since the UV continuum is dominated by radiation from massive stars, $\beta$ is sensitive to the metallicity and age of the stellar population, and especially the dust content within a galaxy \citep[e.g.][]{Cort08, Wilk11}. At high redshifts $z\gtrsim6$, because stellar populations are expected to be uniformly young \citep{Tacc18}, $\beta$ is believed to be mostly affected by dust extinction, making it a useful tracer of chemical enrichment and dust formation in early galaxies. Studies have found that galaxies are generally bluer with increasing redshift \citep{Fink12, Bouw14}, and that fainter galaxies are bluer than brighter ones \citep{Bouw14}. This implies that the dust content in galaxies is decreasing towards earlier times, and that faint galaxies may be less metal enriched than massive ones. 

The sizes and morphologies of galaxies at $z\gtrsim6$ offer another window into galaxy formation and evolution. Sizes are influenced by the host halo's angular momentum \citep[e.g.][]{Mo98}, merger history, and also stellar feedback which can redistribute angular momentum within galaxies~\citep{Brooks11, Genel15, Christensen16}. The luminosity dependence of the galaxy sizes is also affected by the form of the dominating feedback \citep[e.g.][]{Wyit11}. Observations have shown that galaxies at $z\gtrsim6$ are very compact, with physical sizes of order 1~kpc or even smaller \citep[e.g.][]{Oesch10}. The bright galaxies are often found to contain multiple cores, which could be interpreted as merging systems or individual star-forming clumps within a larger structure or disk \citep[e.g.][]{Bowler17}. Observations of the galaxy size--luminosity relation have also put constraints on the specific angular momentum of the disk and the spin parameter of the halo \citep[e.g.][]{Shib15, Kawa18}, providing insights into the angular momentum transfer between disks and halos.

\tmp{Our theoretical picture of galaxy formation has been widely tested against the above mentioned observations of galaxies at $z\gtrsim6$, both by semi-analytic models \citep[e.g.][]{Yung19a, Yung19b} and via cosmological simulations of galaxy formation \citep[e.g.][]{Dave06, Finl06, Finl07, Finl11}.
To obtain the photometric properties of the model galaxies, different groups have used different recipes for dust extinction, combined with stellar population synthesis models. For instance, one can calculate the line-of-sight metal or dust column density \citep[e.g.][]{kli16, cullen17, wilkins17, Ma18a, Voge19, Katz19}, utilize full dust radiative transfer \citep[e.g.][]{narayanan18, behrens18, Ma19, Voge19}, or perform full modelling of dust formation, evolution, and destruction, which has been explored semi-analytically \citep[e.g.][]{popping17, vijayan19}, semi-numerically \citep[e.g.][]{mancini15, mancini16}, and with numerical simulations \citep[e.g.][]{bekki15a, bekki15b, mckinnon17, aoyama17, aoyama18, aoyama20}. While it is desirable to include all details of dust modeling when comparing simulations to observations, the lack of observations at high redshifts also makes it difficult to test the predictions of the dust distribution in simulations. We thus aim to use simple assumptions about dust attenuation to examine the photometric properties of simulated high-$z$ galaxies.% Such predictions from the simulations will also be tested in the near future with next-generation instruments, such as \jwst.
% Since the model parameters are often tuned to math observations at low redshifts \citep[e.g.][]{Voge14, Dubo14, Scha15, Dave16, Spri18, Dave19}, comparisons with high-$z$ observations hint upon the differences in the galaxy formation physics at high and low redshifts.
}

In this work, we study the photometric properties of galaxies at $z\approx6$ from the \simba\ cosmological simulations \citep{Dave19}, a successor to the \mufasa\ project \citep{Dave16}. We examine the UVLF, UV continuum slope, and sizes of the simulated galaxies. \simba\ is calibrated to reproduce the global evolution of the galaxy mass function and quenched galaxy population from $z\sim 6\to 0$, thus providing a plausible feedback model that can be tested against $z\gtrsim 6$ observations. 
%\tmp{\simba\ also includes an on-the-fly dust production, growth, and destruction model, with the dust properties already examined by \citet{Li19}.} 
In addition to investigating the nature of galaxies at the reionization-epoch, we make predictions for future \jwst\ observations. \jwst\ will allow us to probe the rest-frame optical radiation from those $z\sim6$ galaxies, while \hst\ cannot. It thus offers a window for understanding both the rest-UV and rest-optical observations and the link between them. We especially present our mock image generation routine, since the ability of cosmological simulations to resolve structures within galaxies provides a powerful tool to probe the morphologies of high-$z$ galaxies \citep{Ma18b}.

This paper is organized as follows. In Section~\ref{sec:methods} we briefly introduce the simulations and the tools we use to generate mock observations. Section~\ref{sec:results} presents the UVLF, UV continuum slope, and size measurements from the simulated galaxies, and explores predictions for future \jwst\ observation. We summarize our work in Section~\ref{sec:conclusions}.

\section{Simulations and Analysis}
\label{sec:methods}

\subsection{The \simba\ simulations}
\label{sec:sims}

This work uses two new simulations in the \simba cosmological hydrodynamic simulation suite. We refer the readers to \citet{Dave19} for a detailed description of the input physics, and summarize the main points here.  \simba\ uses the {\sc Gizmo} cosmological gravity+hydrodynamics code with its Meshless Finite Mass (MFM) solver \citep{Hopkins15, Hopkins17}.  It includes non-equilibrium radiative cooling from H, He, and metals, and incorporates star formation using an H$_2$-based Schmidt-Kennicutt relation \citep{Kennicutt98}. The H$_2$ fraction computed based on the sub-grid prescription of \citet{Krumholz11}, with a minimum metallicity of 1\% solar.  Star formation driven winds are modeled as kinetic decoupled two-phase outflows, with the mass loading factor scaling with stellar mass as given by \citet{Angles17b}.  Black holes are seeded and grown primarily using a torque-limited accretion model~\citep{Angles17a}, and feedback is in the form of radiative winds, jets, and X-ray momentum.  These modules work together to yield good agreement with a wide range of galaxy~\citep{Dave19} and black hole~\citep{Thomas19} observables across cosmic time.

\tmp{We describe in more detail the metal enrichment, stellar feedback, and dust growth models below, which are the most relevant for our work. The chemical enrichment model tracks eleven elements (H, He, C, N, O, Ne, Mg, Si, S, Ca, Fe) during the simulation, with enrichment tracked from Type II supernovae (SNe) \citep{nomoto06}, Type Ia SNe \citep{iwamoto99}, and Asymptotic Giant Branch (AGB) stars \citep{oppen06}. We note that \simba\ no longer applies an arbitrary reduction of yields by a factor of 2 that was previously needed to match the mass-metallicity relation, and instead lock individual metals into dust. The star formation-driven galactic winds are implemented as decoupled two-phase winds, with 30\% of wind particles ejected with a temperature set by the supernova energy minus the wind kinetic energy. The mass loading factor $\eta$ is a broken power law of stellar mass which follows results from the Feedback In Realistic Environments (FIRE) zoom simulations \citep{Angles17b}, but is capped constant below an $M_*$ corresponding to 16 star particles. $\eta$ is further suppressed at $z>3$. These steps prevent excessive feedback so that poorly-resolved galaxies can grow. The wind velocity scalings follow \citet{muratov15}. \simba\ also implements metal-loaded winds, which extract some metals from nearby particles to represent the local enrichment by the supernovae driving the wind.}

\tmp{Dust growth and destruction processes are tracked on the fly, where dust is passively advected following the gas particles. \simba assumes that dust grains have the same constant radius and density. Dust is produced by condensation of a fraction of metals from SNe and AGB ejecta, following the prescription by \citet{popping17}. Once dust grains are seeded, they grow by accreting gas-phase metals, with an accretion time scale following \citet{hirashita00} and \citet{asano13}. Dust grain erosion follows the approximation of the thermal sputtering rate of grain radii derived by \citet{tsai95}. \simba\ implements a subgrid model for dust destruction by SN shocks \citep{dwek80, seab83, mckee87, mckee89}, and additionally destroy dust completely in hot winds and during star formation and AGN X-ray heating.}

The simulation in \citet{Dave19} had a ($100\ h^{-1}$Mpc)$^3$ volume (\simba-100). Here we run simulations with identical input physics but having volumes of ($25\ h^{-1}$Mpc)$^3$ (denoted as \simba-25) and ($50\ h^{-1}$Mpc)$^3$ (\simba-50).  A minor modification to the \simba-100 mass loading factor scaling is that it is held constant at $M_*<2.9\times 10^8\ M_\odot$ (which is the \simba-100 galaxy mass resolution limit), rather than continuing its power law increase down to each simulation's galaxy mass resolution limit; this was found necessary in order to prevent very large mass loading factors in small early galaxies that over-suppressed early star formation.  Like the large-volume \simba-100 run, these new runs also have $1024^3$ dark matter particles and $1024^3$ gas elements. As such they have mass resolutions $64\times$ and $8\times$ better than the \simba-100, respectively.  This is necessary to better model small early galaxies. Specifically, \simba-25 has a dark matter particle mass resolution of $1.5\times10^6\ M_\odot$ and initial gas element mass resolution of $2.85\times10^5\ M_\odot$, with the corresponding numbers in \simba-50 being 8 times larger. The minimum gravitational softening lengths are $0.125\ h^{-1}$kpc (comoving) and $0.25\ h^{-1}$kpc in \simba-25 and \simba-50, respectively. Notably, \simba-25 resolves halos down to approximately the H cooling limit ($M_{\rm halo}\sim 10^8\ M_\odot$ at $z\sim 6$), so should capture the vast majority of cosmic star formation during the EoR.  \simba-50 enables us to extend the dynamic range and test resolution convergence effects. \simba-25 and \simba-50 are run down to $z\approx 4$, but for this work we only use $z\ge 6$ snapshots.  All runs adopt a \citet{Planck16} cosmology with $\Omega_m = 0.3$, $\Omega_\Lambda = 0.7$, $\Omega_b = 0.048$, $h = 0.68$, $\sigma_8 = 0.82$, and $n_s = 0.97$. 

\subsection{The galaxy sample}

Galaxies are identified in post-processing using the {\sc yt}-based package {\sc Caesar}, which uses a 6-D friends of friends algorithm. We consider galaxies with $\ge32$ star particles. This gives us 13622 galaxies in \simba-25, and 9588 in \simba-50 in the $z=6$ snapshots. For the stellar mass function and the UVLF, we use all galaxies in this sample. For the UV continuum slope and size measurements, we only use galaxies with $\MUV\le-16$ in \simba-25 and those with $\MUV\le-18$ in \simba-50 when assuming the \citet{Calz00} extinction law. These magnitude thresholds are the positions where the UVLFs turn over because the star formation histories cannot be adequately captured with the adopted mass resolutions (see Section~\ref{sec:results} for details). This reduces our sample size to 2699 galaxies in \simba-25, and 2608 in \simba-50 at $z=6$.  We also briefly examine the stellar mass function and the UVLF at $z=8$.

\subsection{Mock photometry, images, and spectra}
\label{sec:pyloser}

We use the {\sc Pyloser} package~\footnote{\tt https://pyloser.readthedocs.io/en/latest/} to create synthetic spectra and images for the simulated galaxies. {\sc Pyloser} treats each star particle in a galaxy as a single stellar population with a given age and metallicity. It generates a mock spectrum for the star particle using the Flexible Stellar Population Synthesis (FSPS) library of population synthesis models \citep{Conr09, Conr10}. We include nebular emission, which is implemented in FSPS by a pre-computed {\sc Cloudy} table, implemented by \citet{byler17}. By default, we use the MIST isochrones \citep{mist1, mist2, mist3, mist4, mist5} and MILES spectral library \citep{miles1, miles2} and assume a \citet{Chab03} initial mass function. In Appendix~\ref{sec:compare_bpass}, we will show that using different isochrone models, especially one that includes stellar binaries, yields very similar UVLF and $\beta-\MUV$ relation as using MIST. This is consistent with the findings of \citet{Choi17} that the predictions on the far-UV spectrum by different isochrone models agree well with each other.

To take into account dust extinction, we first calculate the metal column density and the mass-weighted average metallicity along the line of sight to each star particle. \tmp{The metal column density is converted to the total extinction in the $V$ band, $A_V$, via the $A_V$--metal column density relation in the Milky Way \citep{Wats11}. We then scale the $A_V$ value by the ratio of the dust-to-metal ratio (DTM) of the sightline to that of the Milky Way \citep{Dwek98, Wats12}, where the DTM is calculated using the average metallicity of the sightline and a fitting formula to the simulated DTM--gas-phase metallicity relation of \simba-100 \citep{Li19}.
We note that in principle we could compute the dust column density directly from the \simba\ outputs. We refrain from doing so because it is unclear how well the locally-calibrated dust model works at $z\ga6$, owing to lack of observations and a lack of constraints on the input model parameters. There is also no straightforward way to convert the simulated dust distribution into amounts of extinction, without full dust radiative transfer. The $A_V$--metal column density relation is, in contrast, better constrained by observations. We therefore defer a more careful investigation of the dust model at these redshifts to future work, and continue with the approximation that dust traces metals.}

Given an $A_V$ value for each star particle, we then extinct its spectrum by assuming an extinction law. We primarily use the \citet{Calz00} law, but we also consider the extinction law from \citet{Salm16}, which multiplies the \citet{Calz00} law by $(\lambda/\lambda_V)^\delta$, where $\delta = 0.62 \log E(B-V) + 0.26$ and $\lambda_V=5500$~{\AA}. Such a form lets galaxies with high color excess have a shallower, starburst-like law, and those with low color excess have a steeper, SMC-like law. Finally, we also examine the SMC extinction law~\citep{Gord04}. For creating the mock images, we only use the \citet{Calz00} law since it yields the best match of the simulated UVLF and $\beta-\MUV$ relation to observations (see below and Section~\ref{sec:results}).

For each {\sc Caesar}-identified galaxy, {\sc Pyloser} sums the spectra of all the star particles in it and convolves the summed spectra with desired filters to get the AB magnitudes. We use the rest-frame spectra to compute the absolute UV magnitudes, and the redshifted spectra to calculate the apparent magnitudes for relevant \hst\ and \jwst\ bands. The magnitude at rest-frame 1500 {\AA}, $\MUV$, is calculated by the convolving the rest-frame spectra using a boxcar filter centered at 1500 {\AA} with 400 {\AA} in width. For the \jwst\ bands, we especially focus on \jwst\ F115W and F444W, which correspond to rest-frame 1600 {\AA} and 6300 {\AA} at $z=6$.  {\sc Pyloser} is also able to make mock images by binning the star particles into pixels, and analogously compute the magnitudes in each pixel. We use the pixel sizes of the corresponding \hst\ and \jwst\ bands to generate the images. Specifically, F115W and F444W have pixels scales of $0.031$ and $0.063$~arcsec respectively, which correspond to $0.18$ and $0.37$ physical kpc at $z=6$.

\subsection{An example galaxy}

To illustrate the sort of outputs obtainable from {\sc Pyloser}, we show images and spectra for a selected $z=6$ galaxy from \simba-25. This galaxy has a stellar mass of $6\times10^9\ M_\odot$, a SFR of $24\ M_\odot$/yr, and a gas-phase SFR-weighted metallicity of $0.28\ Z_\odot$.

\begin{figure*}
	\includegraphics[width=\linewidth]{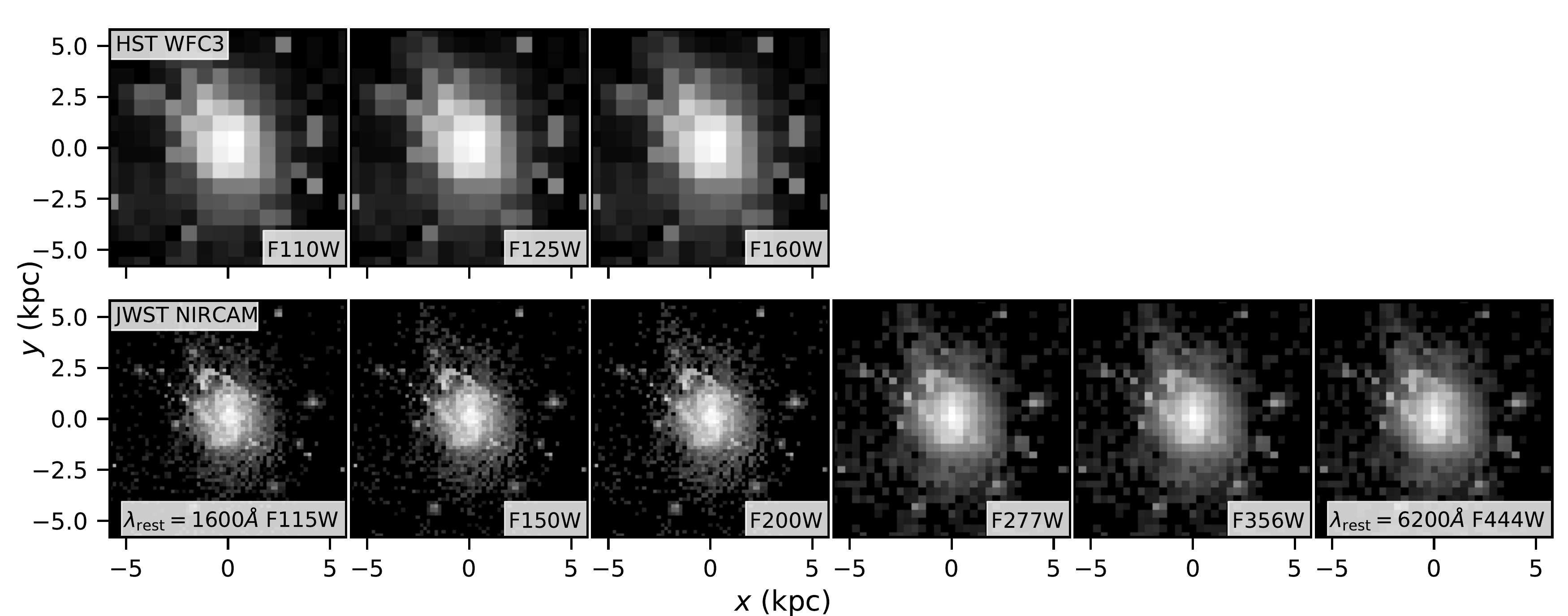}
    \caption{An example of the mock images of a galaxy in \simba-25 at $z=6$ created with {\sc Pyloser} assuming the \protect\citet{Calz00} extinction law. This galaxy has a stellar mass of $6\times10^9\ M_\odot$, a SFR of $24\ M_\odot$/yr, a gas-phase SFR-weighted metallicity of $0.28\ Z_\odot$, and $\MUV = -20.88$. The $x,y$ coordinates are in units of physical kpc. The images have a side length of four times the 3D stellar radius of the galaxy. Top and bottom panels show images in three \hst\ IR bands and six \jwst\ bands, created using the corresponding pixel scales. The wavelengths increase from left to right. The \jwst\ F115W and F444W bands correspond to rest-frame 1600 {\AA} and 6300 {\AA}, respectively. The images are not convolved with PSFs and no noise has been added.}
    \label{fig:example_images}
\end{figure*}

Figure~\ref{fig:example_images} shows example images generated by {\sc Pyloser} for the selected galaxy in \simba-25 at $z=6$ in various \hst\ WFC3 and \jwst\ NIRCAM bands. The $x,y$ coordinates are in units of physical kpc. The images are shown out to twice the 3D stellar radius of the galaxy, which is the largest distance of a star particle to the galaxy's center of mass as determined by {\sc Caesar}. From left to right, the effective wavelengths of the filters increase. Each image is created using the corresponding pixel scale of the instrument/filter. The images are not convolved with PSFs and no noise is added. In this optimal setup, \jwst\ clearly resolves more detailed structures of the galaxy than \hst, such as clumps, streams, and the central core. The \hst\ WFC3 filters have a pixel scale of $0.13$~arcsec, which is about four times and twice as large as the pixel scales of \jwst\ F115W and F444W, respectively. Similarly, the \jwst\ F115W image shows more structures than the F444W one. Interestingly, this galaxy has very similar morphologies in the rest-frame UV (e.g. F115W) and rest-frame optical (e.g. F444W) bands. In fact, we find that the simulated galaxies exhibit a general trend of having similar morphologies in rest-UV and rest-optical, owing to young ages of the stellar populations (see Section~\ref{sec:size} for details).

\begin{figure}
\includegraphics[width=\columnwidth]{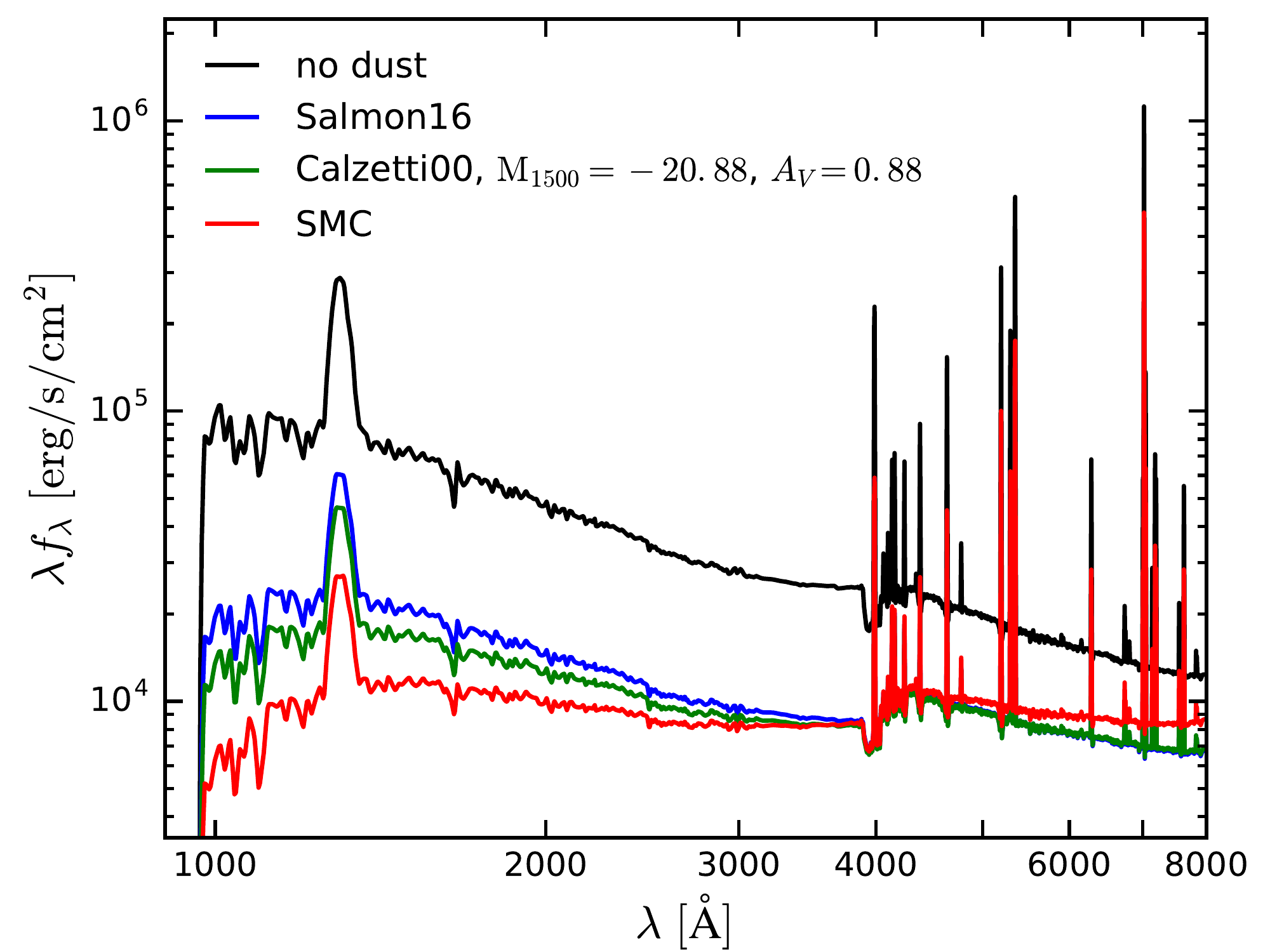}
    \caption{Mock spectra of the same galaxy as shown in Figure~\ref{fig:example_images} at $z=6$, assuming no dust (black line), the \protect\citet{Salm16} (blue), \protect\citet{Calz00} (green), and SMC (red) extinction laws. When using the \protect\citet{Calz00} extinction law, this galaxy has $\MUV=-20.88$ and $A_V=0.88$.}
    \label{fig:example_spectra}
\end{figure}

Figure~\ref{fig:example_spectra} presents an example of mock spectra generated for the selected galaxy in \simba-25. The black line shows the extinction-free spectrum, and the blue, green, and red lines represent the attenuated spectrum assuming the \citet{Salm16}, \citet{Calz00}, and SMC laws, respectively. Assuming the \citet{Calz00} law, this galaxy has a UV magnitude of $\MUV=-20.88$ and $A_V=0.88$\footnote{\tmp{$A_V$ is computed by subtracting the extincted galaxy spectrum from the extinction-free one, and convolving this with the Johnson $V$ filter.  Hence it depends on the assumed extinction law.  Different extinction laws results in $\lesssim0.3$ differences in the $A_V$ values.}}. Hence this is a galaxy that is typically detectable with \hst\ at these epochs. The \citet{Salm16} law exhibits the least extinction, while the SMC law generates the most dust attenuation. The galaxy displays a very blue continuum and strong emission lines characteristic of early galaxies, but also shows a noticeable 4000\AA\ break indicating an underlying older stellar population.  

\section{Main Results}
\label{sec:results}

\subsection{The stellar mass function and the UVLF}
\label{sec:UVLF}

\begin{figure}
	\includegraphics[width=\columnwidth]{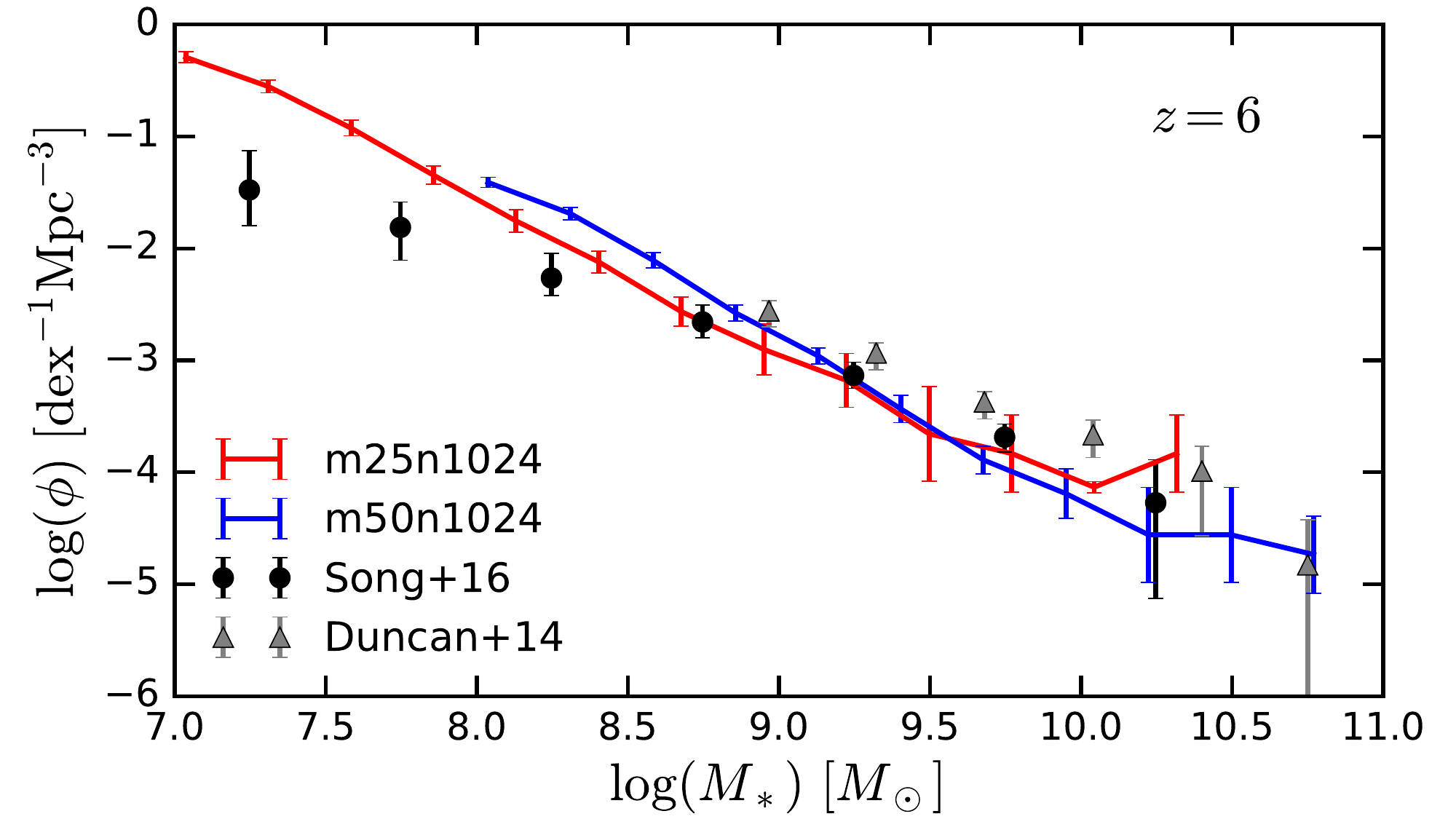}
	\includegraphics[width=\columnwidth]{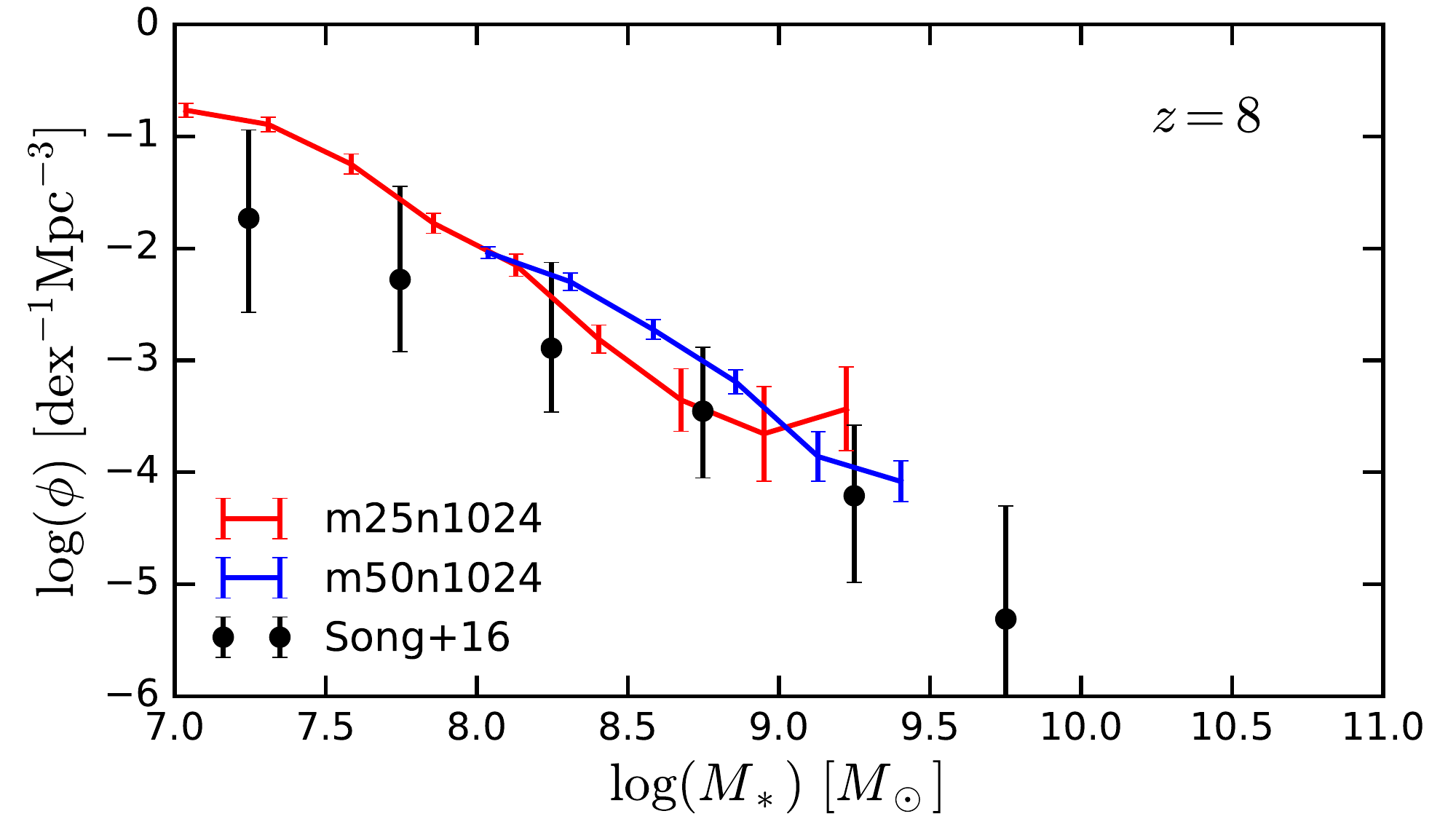}
    \caption{The galaxy stellar mass functions at $z=6$ (top panel) and $z=8$ (bottom panel). Red and blue lines represent the mass functions in \simba-25 and \simba-50, respectively. Errorbars show a jackknife estimate of the error. Circles and triangles illustrate the observational data in \protect\citet{Song16} and \protect\citet{Dunc14}, respectively. The two simulations roughly match the observational data at both redshifts.}
    \label{fig:stellarMF}
\end{figure}

Figure~\ref{fig:stellarMF} shows the stellar mass functions in \simba-25 (red lines) and \simba-50 (blue lines) at $z=6$ (top panel) and $z=8$ (bottom panel). Errorbars represent a jackknife estimate of the error. Circles and triangles represent the observational data in \citet{Song16} and \citet{Dunc14}, respectively. 

The two simulations are in fairly good agreement with the observational data at both redshifts.  Also, the good agreement between the two simulations in the overlapping mass range shows that they are reasonably well converged in mass growth.  \simba-25 overshoots the number of low-mass galaxies at $z=6$. While systematic uncertainties in the observations are generally larger here, this may indicate that \simba\ does not suppress very small galaxy formation sufficiently. \tmp{We note that we suppressed the mass loading factor in low-mass galaxies at high redshift in order to achieve agreement in the SFR function of low-mass galaxies, and consequently in the UVLF that we will discuss in the next section, but this appears to still slightly overproduce the stellar mass in these galaxies.}
% Overall, \simba\ grows stellar mass in early galaxies in good agreement with observations.

\begin{figure}
	\includegraphics[width=\columnwidth]{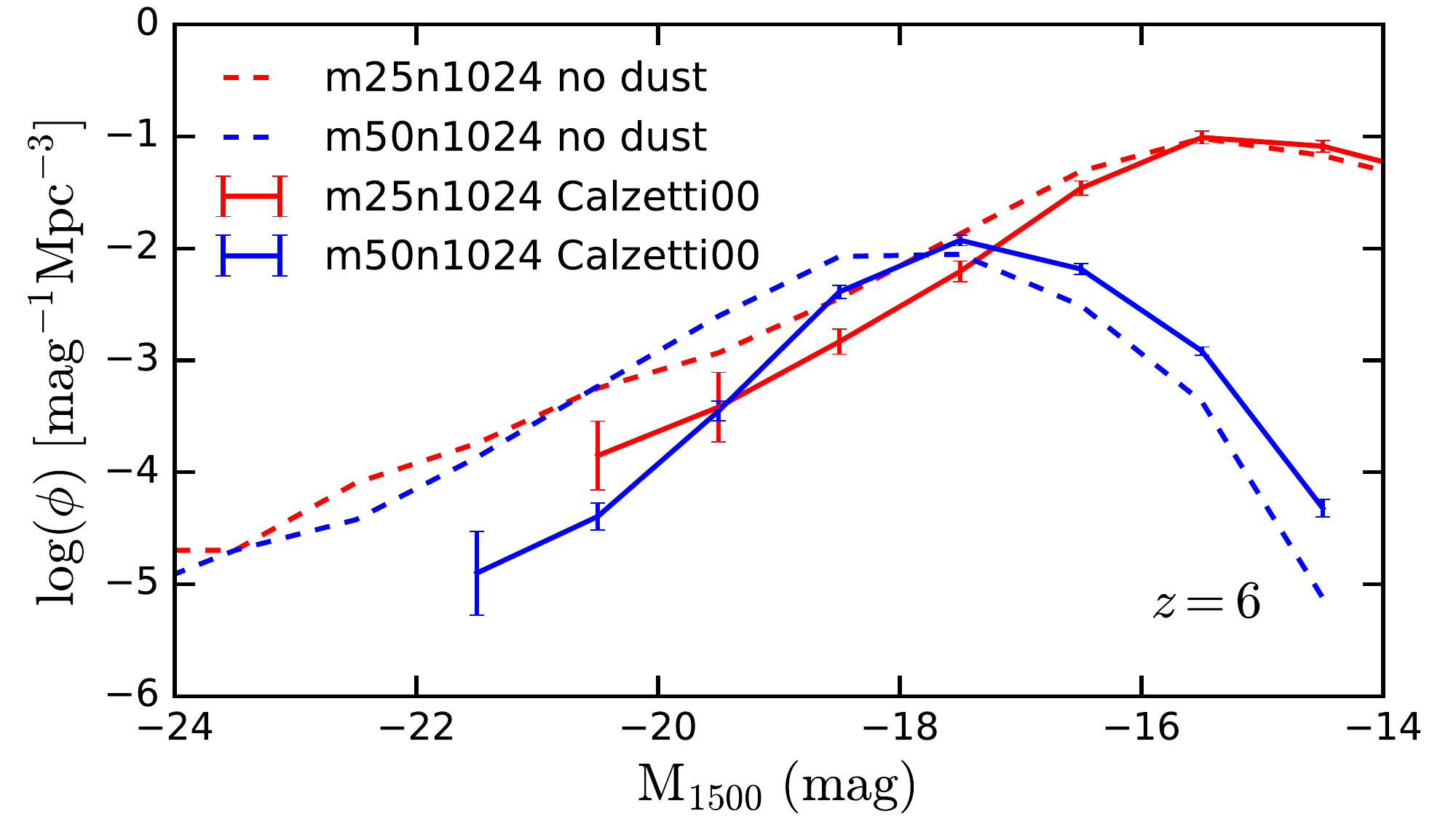}
	\includegraphics[width=\columnwidth]{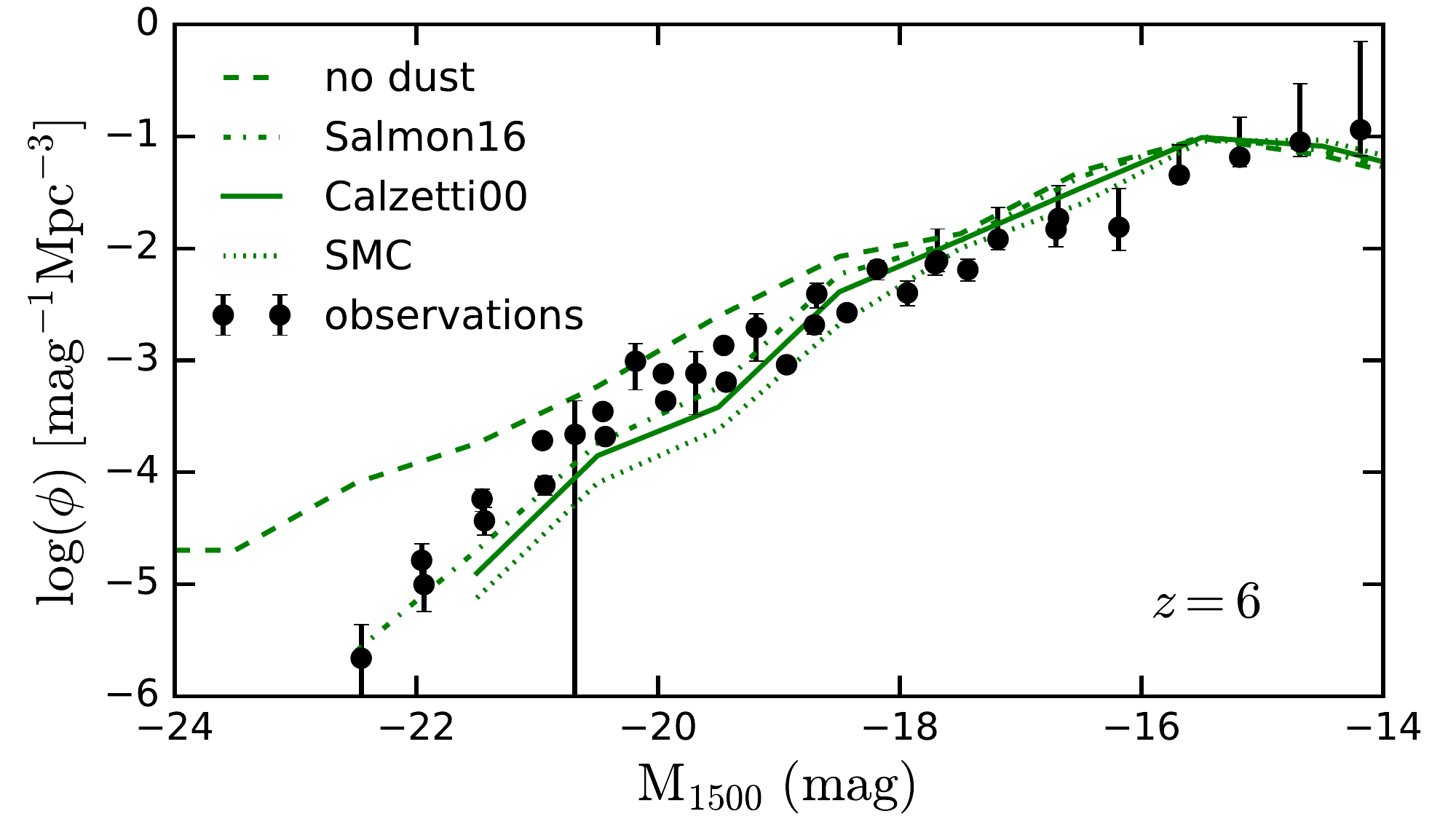}
	\includegraphics[width=\columnwidth]{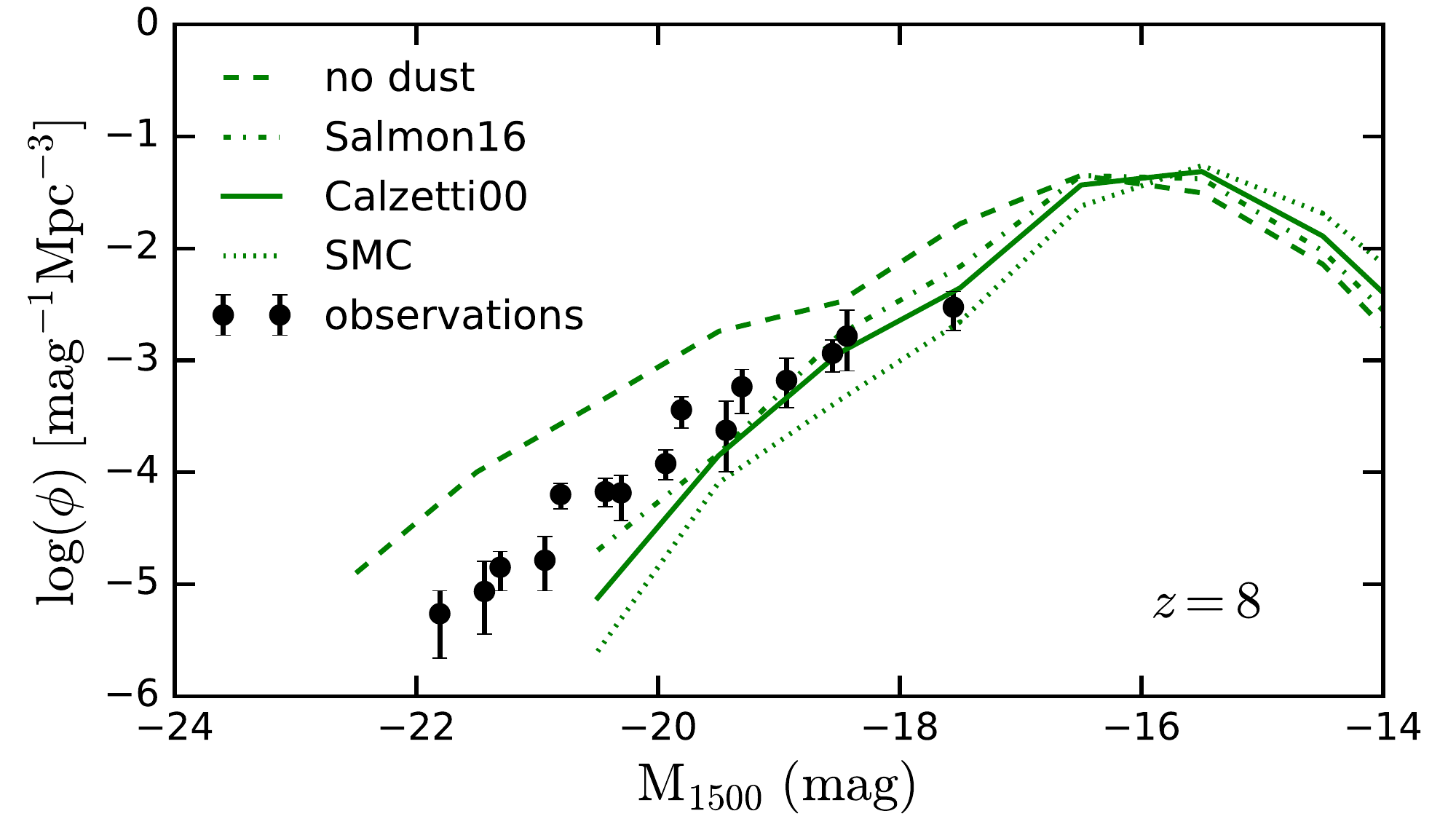}
    \caption{Top: the $z=6$ UV luminosity functions in \simba-25 (red) and \simba-50 (blue). Solid lines represent UVLFs using the \protect\citet{Calz00} extinction law, and dashed lines show the no dust LFs. Errorbars represent a jackknife estimate of the error and are only plotted for the dust-attenuated UVLFs. Middle: combined $z=6$ LFs of \simba-25 and \simba-50. Green dashed, solid, dot-dashed, and dotted lines show LFs assuming no dust, \protect\citet{Calz00} law, \protect\citet{Salm16} law, and SMC law, respectively. Black circles represent the observational data compiled from \protect\citet{Bouw15}, \protect\citet{Fink15}, and \protect\citet{Bouw17}. Bottom: similar to the middle panel, but showing results at $z=8$. Observational data comes from \protect\citet{Bouw15} and \protect\citet{Fink15}. The UVLF is not very sensitive to the dust extinction law. The simulations roughly match the observations at $z=6$, but underpredict the number of bright galaxies at $z=8$ with dust-attenuation.}
    \label{fig:UVLF}
\end{figure}

Figure~\ref{fig:UVLF} presents the simulated rest-frame 1500 {\AA} LFs at $z=6$ (top and middle panels) and $z=8$ (bottom panel). Here we use all galaxies in our sample, with no stellar mass limit, in order to assess resolution convergence. Since these small young objects are very gas-rich, they typically have $\ga 100$ gas and star particles in total, which is generally regarded as sufficient to be well-resolved.  The red and blue lines in the top panel show the $z=6$ UVLFs in \simba-25 and \simba-50, respectively. Solid and dashed lines represent UVLFs assuming no dust and the \citet{Calz00} extinction law, respectively. Errorbars illustrating the jackknife estimate of the error are only plotted for the dust-attenuated UVLFs. 

The UVLFs in the two simulations are reasonably well converged in the $\MUV$ range where they overlap. The \simba-25 UVLF shows a turnover at $\MUV=-16$, and \simba-50 at $\MUV=-18$. These turnovers are not physical. They occur because the mass resolutions of these simulations can no longer adequately capture the star formation histories of the low-mass galaxies, owing to our stochastic approach to star formation which occurs quite rarely in such small systems with high mass loading factors. Therefore in the analysis on the UV continuum slope and size measurements below, we only focus on galaxies with $\MUV\le-16$ in \simba-25, and those with $\MUV\le-18$ in \simba-50.

We combine UVLFs in \simba-25 and \simba-50 by taking the value of the combined LF in each $\MUV$ bin to be the higher one of the LFs in the two simulations~\citep[following][]{Dave06}\footnote{\tmp{We note that our approach of combining the LFs tends to bias the combined LF high. One might prefer an weighted average, but there is no obvious way to determine the weights, since one needs to define which simulation is ``better'' and meanwhile take into account cosmic variance. Taking the maximum value of each bin is well-defined, since the small box simulation is incomplete at the bright end and the large box is incomplete at the faint end. We have also verified that taking a number-squared weighted-average of the LFs \citep[e.g.][]{Voge19} yields very similar results as taking the maximum value.}}. The middle panel of Figure~\ref{fig:UVLF} illustrates the combined UVLFs from \simba-25 and \simba-50 at $z=6$. The green dashed, dot-dashed, solid, and dotted lines represent the combined UVLFs assuming no dust, the \citet{Salm16} law, \citet{Calz00} law, and the SMC law, respectively. The observational data shown by black circles is a compilation of the data in \citet{Bouw15}, \citet{Fink15}, and \citet{Bouw17}. 

The no-dust UVLF clearly overshoots the number of bright galaxies, showing that despite the UVLF being at face value consistent with a power law, in fact the brightest UV galaxies at these epochs must already be substantially dust attenuated. All three extinction laws yield simulated UVLFs that are in agreement with the observations. The SMC law produces the most dust-attenuation, so the bright end of the LF is suppressed a bit more. Similarly, the \citet{Salm16} law gives more bright galaxies, but the differences produced by the three extinction laws are too small to be distinguishable with the observational data. This indicates that current measures of the UVLF are not strongly sensitive to the dust attenuation law assumed, for reasonable choices. \tmp{An agreement between the simulated dust-attenuated UVLF and the observations has also been found in a number of previous works using different assumptions on dust properties and post-processing techniques \citep[e.g.][]{kli16, mancini16, wilkins17, Ma18a, Ma19}. While the effect of dust attenuation is already observable at $\MUV<-18$ in \simba, other works seem to show that dust extinction starts becoming important for $\MUV<-20$ at such high redshift. This indicates that the \simba\ galaxies have a bit more dust extinction at $z=6$. As we will show below, this effect is more prominent at $z=8$.}

The bottom panel of Figure~\ref{fig:UVLF} is similar to the middle panel, but showing results at $z=8$. Interestingly, the no-dust UVLF still overshoots the observations of \citet{Bouw15} and \citet{Fink15}, but all the dust-attenuated UVLFs now under-predict the number of galaxies with $\MUV<-18$. One possibility is that the assumed normal stellar populations and IMF used to generate the UV emission is reasonable at $z\sim 6$, but at $z\sim 8$ they under-predict the UV flux.  Another possibility is that there is too much dust attenuation in the simulations at these epochs.  The latter seems more likely given the large difference between the no-dust and dust-attenuated UVLFs at $z=8$ compared to $z=6$. \tmp{We will examine the physical reason for this excess dust extinction in Section~\ref{sec:metallicity}.}

\begin{table}
 \caption{\tmp{Faint-end slopes of the simulated UVLFs at $z=6$ (top row) and 8 (bottom row), fitted using galaxies with $-19<\MUV<-16$. Middle and right columns list the slopes without dust and with the \citet{Calz00} extinction law respectively.}}
 \label{tab:faint_end_slope}
 \begin{tabular}{ccc}
  \hline
  \tmp{Redshift} & \tmp{Slope without dust} & \tmp{Slope with Calzetti law} \\
  \hline
  \tmp{6} & \tmp{-1.6} & \tmp{-1.7} \\
  \tmp{8} & \tmp{-1.7} & \tmp{-1.9} \\
  \hline
 \end{tabular}
\end{table}

\tmp{Overall, the \simba\ stellar mass function and UVLF are in descent agreement with observations at $z=6$ and 8, albeit the $z=8$ UVLF being a bit low. Before moving on to the UV continuum slope, we briefly comment on the faint-end slope of the UVLF. Table~\ref{tab:faint_end_slope} lists the faint-end slopes of the simulated UVLFs at $z=6$ (top row) and 8 (bottom row), fitted using galaxies with $-19<\MUV<-16$. Middle and right columns list the slopes without dust and with the \citet{Calz00} extinction law respectively. Generally, we find values shallower than $-2$, which may be slightly inconsistent with observations \citep[e.g.][]{Bouw15, Fink15, Bouw17}. However, we have not attempted to estimate uncertainties in simulations, which come from both resolution effects and sub-grid models. Caution therefore needs to be taken when interpreting the faint-end slopes of the simulated UVLF.}

\subsection{The UV Continuum Slope}
\label{sec:beta}

\begin{figure}
	\includegraphics[width=\columnwidth]{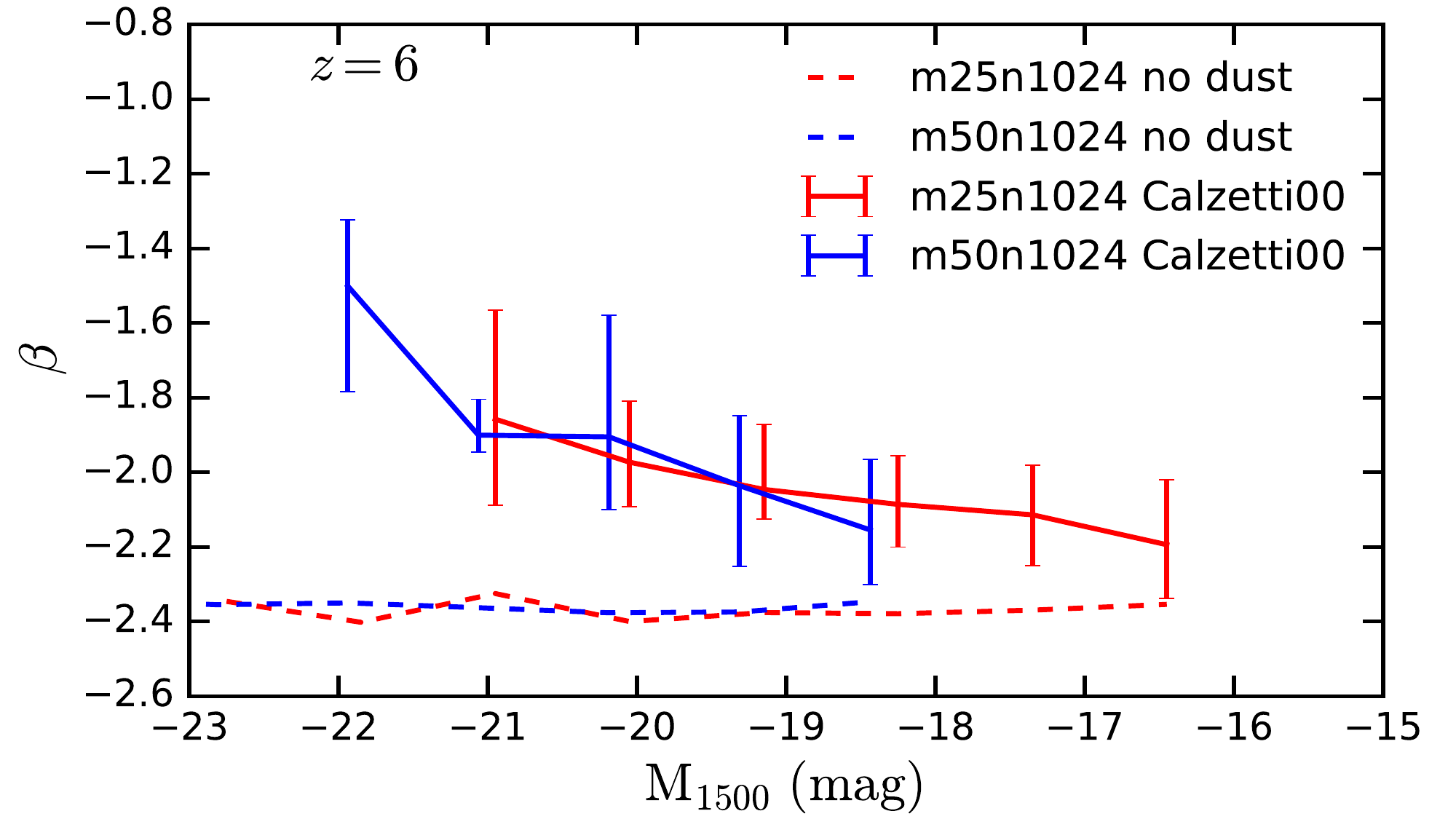}
	\includegraphics[width=\columnwidth]{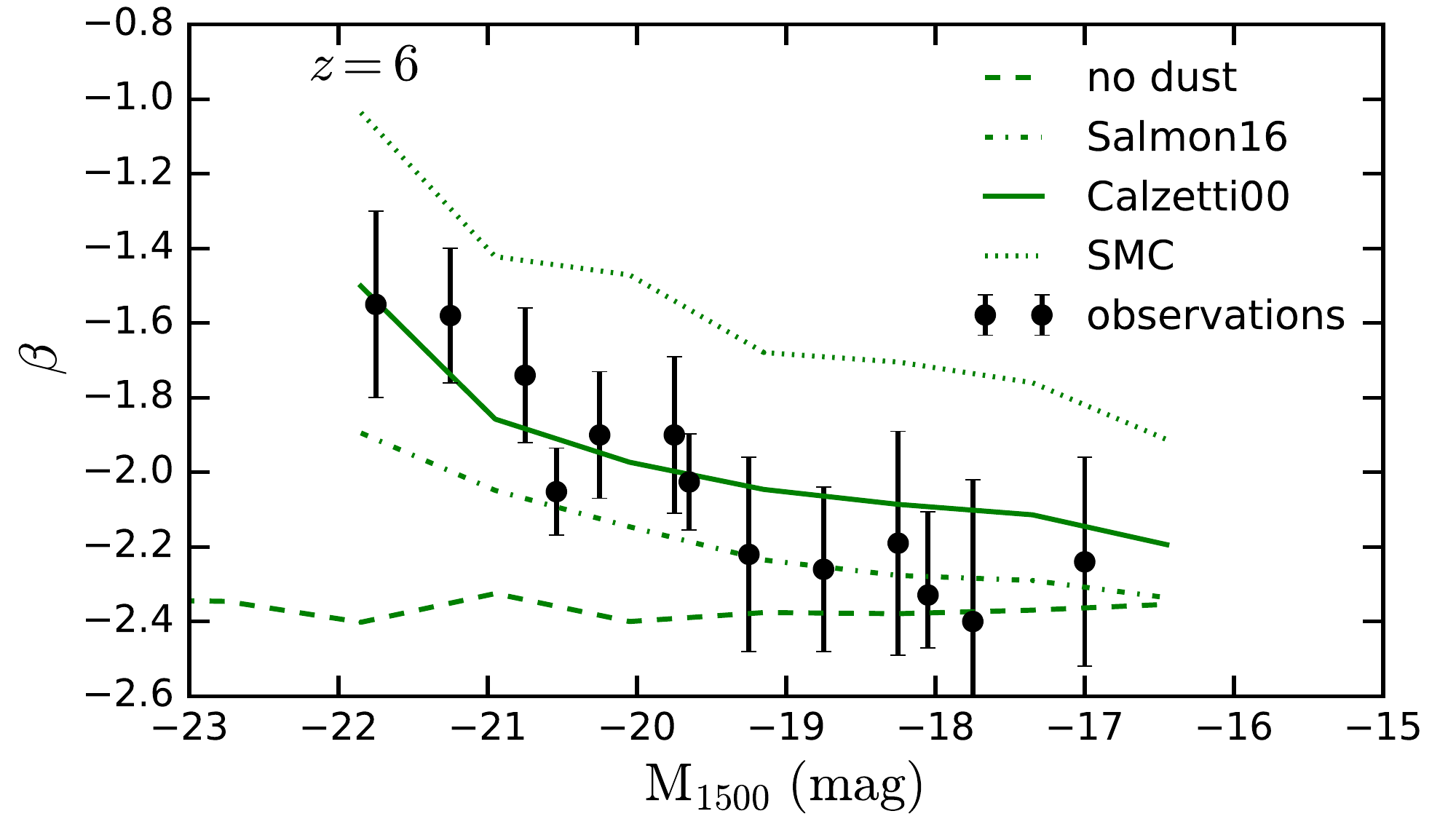}
    \caption{Top: the $z=6$ UV continuum slope ($\beta$) -- UV luminosity ($\MUV$) relations in \simba-25 (red) and \simba-50 (blue). Solid and dashed lines represent the $\beta-\MUV$ relations assuming the \protect\citet{Calz00} extinction law and no dust respectively. Errorbars illustrate $1\sigma$ spread in $\beta$ and are only shown for the dust-attenuated $\beta-\MUV$ relations. Bottom: combined $\beta-\MUV$ relations of \simba-25 and \simba-50. The green dashed, solid,  dot-dashed, and dotted lines represent the relations assuming no dust, \protect\citet{Calz00}, \protect\citet{Salm16}, and SMC laws respectively. Black circles showing the observational data are compiled from \protect\citet{Bouw14} and \protect\citet{Fink12}. The $\beta-\MUV$ relation is sensitive to the dust attenuation law, and the \protect\citet{Calz00} law gives the best fit to the observations.}
    \label{fig:beta}
\end{figure}

A more stringent test of dust attenuation is provided by examining the UV continuum slope $\beta$.  We compute $\beta$ by fitting a power-law slope using the three rest-frame magnitudes at rest-frame 1500 {\AA}, 2300 {\AA}, and 2800 {\AA}, which roughly mimics the observational procedure. We have verified that fitting a power-law to the rest-frame spectra directly yields very similar results. 

Figure~\ref{fig:beta} shows the simulated $\beta-\MUV$ relations at $z=6$. Colors and linestyles are the same as in Figure~\ref{fig:UVLF}. The top panel compares the results from \simba-25 (red) and \simba-50 (blue), without dust attenuation (dashed) and assuming \citet{Calz00} law (solid). Errorbars represent $1\sigma$ spread in $\beta$ within the galaxy sample. As motivated in Section~\ref{sec:UVLF}, we only use galaxies with $\MUV\le-16$ in \simba-25 and those with $\MUV\le-18$ in \simba-50. The two simulations are well-converged in the $\MUV$ range where they overlap.

Without dust attenuation, the $\beta-\MUV$ relations are flat, indicating that all galaxies have similar ages and metallicities. This is because the star formation rate--stellar mass relation in the simulations at $z\approx6-8$ has a slope of $\approx1$. Thus all galaxies require similar amounts of time to assemble their stellar masses, leading to roughly constant age with stellar mass \citep[see also][]{Tacc18}. \tmp{This flat no-dust $\beta$ is consistent with the findings of \citet{kli16, mancini16}. Other simulations, e.g. \citet{wilkins17, mancini16, Ma18a}, also found either a flat specific star formation rate or a flat mass-weighted age across different stellar masses, confirming our explanation.} Hence the trend in the simulated $\beta$ mostly arises from the differences in dust attenuation in the galaxies.
In reality, $\beta$ is also a function of the uncertain metallicity of stellar population and the star formation history. Changes in these two quantities can lead to a change in $\beta$ by up to $<0.3$ \citep{Wilk11, Tacc18a}. This is much smaller than a change in $\beta$ induced by a change in $A_V$ or the dust extinction law, which we find to be at $\Delta\beta\sim0.8$ level (see analysis below). Therefore $\beta$ is most sensitive to the dust attenuation and can be a good tracer of the dust extinction law.

The bottom panel of Figure~\ref{fig:beta} combines the samples from \simba-25 and \simba-50 to show $\beta-\MUV$, illustrating the effects of our three different extinction laws: \citet{Salm16} (dot-dashed), \citet{Calz00} (solid), SMC (dotted), and no dust (dashed). We compare to observational data of \citet{Bouw14} and \citet{Fink12} shown as the black circles.
\tmp{Overall, \simba\ predicts more dust attenuation in brighter galaxies.  This broadly agrees with \citet{Ma19} who showed that a heavy attenuation is required to reproduce the observed bright-end UVLFs at $z\ge 5$.}

The SMC law, which reddens the galaxy spectra the most (see Figure~\ref{fig:example_spectra}), produces the shallowest UV continuum slopes and thus a $\beta-\MUV$ relation that is too high ($\Delta\beta\sim0.4$) compared to the observations. In contrast, the \citet{Salm16} law has the least extinction and hence generates bluer galaxies with the lowest $\beta-\MUV$ relation. It roughly matches the observational data at $\MUV>-20$, but makes the bright galaxies slightly too blue compared to the observations ($\Delta\beta\sim-0.3$). The \citet{Calz00} law yields the overall best match of the simulations to the observations, though it would be even slightly better if faint galaxies had a flatter attenuation law  like that of \citet{Salm16}.  The best-fitting law being \citet{Calz00} may be in some tension with observations such as \citet{Salm16} and \citet{Reddy16} which tend to show less far-UV attenuation than \citet{Calz00}.  We note that there is some degeneracy between $\beta$ and the intrinsic far-UV emission slope, which may be impacted by different stellar populations such as more binaries or Population III stars.  Given such uncertainties, the broad agreement of \simba\ with observations assuming a \citet{Calz00} attenuation law is preliminarily encouraging. \tmp{Our finding that the simulations favor the \citet{Calz00} law against the SMC law is also consistent with the simulation work of \citet{cullen17}, though \citet{mancini16} found that an SMC-like extinction law with a clumpy dust distribution is preferred. A full exploration of the dust extinction law at $z\sim6$ thus requires dust radiative transfer, which we will defer to future work.}

\begin{figure}
	\includegraphics[width=\columnwidth]{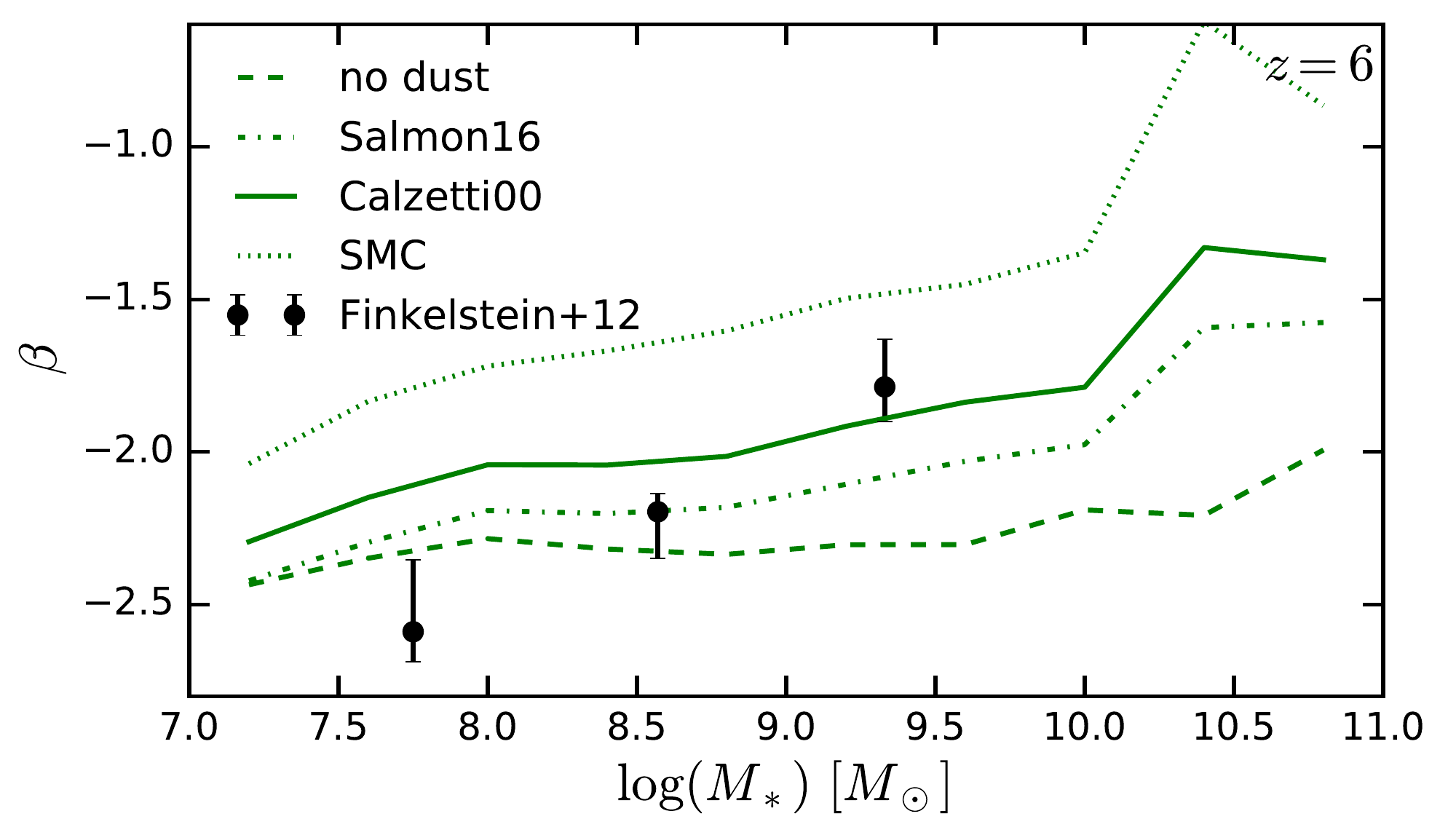}
    \caption{\tmp{Combined $\beta-M_*$ relations of \simba-25 and \simba-50 at $z=6$. The green dashed, solid, dot-dashed, and dotted lines represent the relations assuming no dust, \protect\citet{Calz00}, \protect\citet{Salm16}, and SMC laws respectively. Black circles show the observational data from \protect\citet{Fink12}. The observations indicate much steeper changes of $\beta$ with stellar mass.}}
    \label{fig:beta-Mstar}
\end{figure}

\tmp{Finally, we examine the dependence of $\beta$ on stellar mass, since \citet{Fink12} found a more significant correlation of $\beta$ with $M_*$, than with $\MUV$. Figure~\ref{fig:beta-Mstar} shows the simulated $\beta-M_*$ relations at $z=6$, combined from \simba-25 and \simba-50. Line styles are the same as in the bottom panel of Figure~
\ref{fig:beta}. Since $\MUV$ and $M_*$ are tightly correlated, the overall trend of the simulated $\beta-M_*$ relation is expected not to change much from that of $\beta-\MUV$. However, the observations of \citet{Fink12} clearly show a steeper $\beta-M_*$ relation than our simulations. \citet{kli16} also found a tighter correlation between $\beta$ and $M_*$ in their simulations, but we note that the correlation only gets steep for $M_*<10^7\ M_\odot$, so our findings roughly agree with theirs. This indicates a possible tension between observations and simulations. One issue could be an observational selection effect that the lowest mass galaxies are more visible if they are younger and less dust attenuated, which may bias towards lower $\beta$ values. Future \jwst\ observations will be able to lower the uncertainties in determining both $\beta$ and $M_*$, thus providing stronger constraints on the correlation.}

In summary, \simba\ produces a $\beta-\MUV$ relation in good agreement at $z=6$, assuming the \citet{Calz00} extinction law. We also examined this at $z=8$, but the dynamic range is small and the uncertainties are large owing to the small number of galaxies in both observations and \simba, so this was not particularly enlightening.  IR spectroscopy with \jwst\ to more directly constrain the typical extinction curve in $z\ga 6$ systems will prove quite valuable at discriminating between models of early galaxy and dust growth, and thus potentially provide indirect constraints on mass outflow rates.

\subsection{The metal enrichment of reionization-epoch galaxies}
\label{sec:metallicity}

\begin{figure}
	\includegraphics[width=\columnwidth]{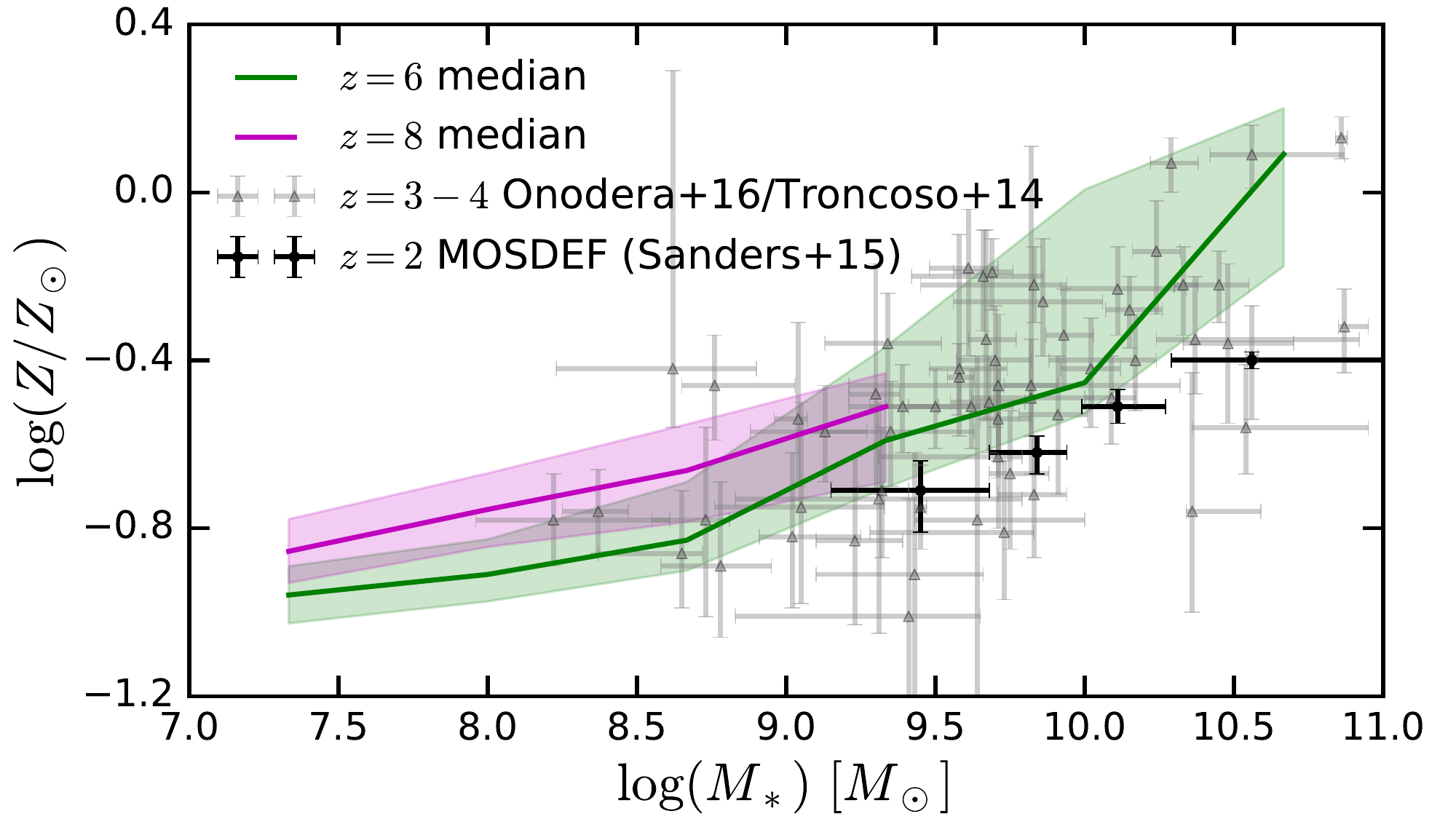}
	\includegraphics[width=\columnwidth]{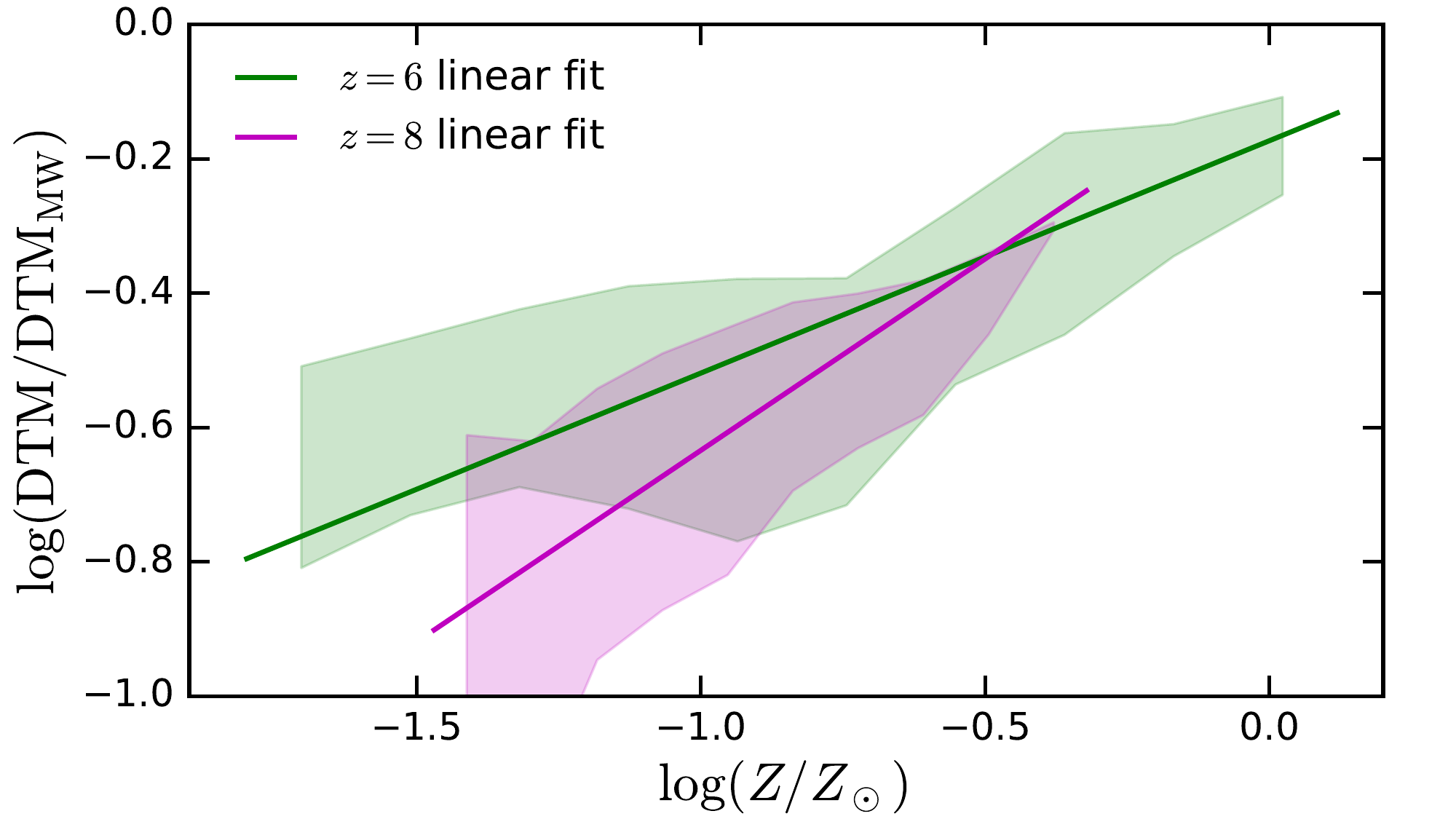}
    \caption{Top: gas-phase mass--metallicity relations in \simba-25 and \simba-50 at $z=6$ (green) and $z=8$ (magenta). Solid lines show the median gas-phase mass--metallicity relations, and shaded regions represent $1\sigma$ spread in the metallicity. Black dots with errorbars show the observational data from the MOSDEF survey at $z=2$ \protect\citep{Sand15}, and gray triangles represent the $z=3-4$ observations of \protect\citet{Onod16} and \protect\citet{Tron14}. \tmp{Bottom: simulated dust-to-metal ratio--gas-phase metallicity relations at $z=8$ and 6. Shaded regions show $1\sigma$ spread in DTM, and solid lines represent power-law fits which are used to calculate the UVLF with {\sc Pyloser}. \simba\ predicts early metal enrichment and high metallicities that would lead to too much dust extinction, which is partly mitigated by a steep DTM that reduces the $A_V$ values of faint galaxies.}}
    \label{fig:MZR}
\end{figure}

\tmp{We now examine the origin of the dust-attenuated $z=8$ UVLF being low compared to observations, as found in Section~\ref{sec:UVLF}. Since we apply an attenuation based on the line-of-sight gas-phase metal column density to each star and modify the $A_V$ values using the simulated DTM--gas-phase metallicity relation, our finding hints upon possible early metal enrichment in the simulations that leads to too high $A_V$ values. Figure~\ref{fig:MZR} examines the gas-phase mass-metallicity relation (MZR, top panel) and the DTM--metallicity scaling (bottom panel) in \simba\ at $z=6$ (green) and $z=8$ (magenta).} \simba-25 and \simba-50 galaxies have been combined to produce the overall relation; they agree well in the overlap region. \tmp{The shaded regions illustrate $1\sigma$ spread in metallicity and DTM. The solid lines represent the median metallicity and linear fits to the DTM-Z relations in the top and bottom panels respectively.} There are no observations yet of the MZR at $z=6-8$, but for comparison we show observations at $z\approx 2$ from the MOSDEF survey~\citep{Sand15} and those at $z=3-4$ by \citet{Tron14} and \citet{Onod16}.  We note that \simba-100 well reproduces the MZR at $z\sim 2$~\citep{Dave19}, and we would expect \simba-25 and \simba-50 to do so as well if they were run down to that redshift.

The MZR predicted in \simba-25 and \simba-50 is consistent with the $z=3-4$ observations, although the observations have large scatter and the inferred metallicity depends strongly on the assumed strong-line calibration. Remarkably, the simulated MZR is higher at $z\sim 6-8$ than at $z=2$.  This may seem counter-intuitive at first, because metallicity is often regarded as a cumulative record of past star formation.  However, as \citet{Finl08, Dave12} and others have argued, in fact this is not correct when the inflow rates are large.  Instead, the MZR reflects a recent balance between inflows and outflows, approximately given by $Z\approx y\dot{M}_{\rm in}/(1+\eta)$, where $y$ is the yield, $\dot{M}_{\rm in}$ is the accretion rate onto the ISM, and $\eta$ is the mass loading factor.  The mass dependence and evolution of the MZR thus reflects that of the inflow rate and mass loading factor.

In \simba, we attenuate the high-$z$ mass loading factor in small galaxies ($M_*<2.9\times 10^8\ M_\odot$) by forcing a constant mass loading factor below this mass. This was found to be necessary to generate early galaxy growth as observed; if we extrapolated our mass-dependent $\eta$ to very small systems, we end up with extremely large values of $\eta$ that prevent any early growth at all.  With this choice, we are able to broadly reproduce observed early stellar mass growth.

However, this choice also means that the MZR has a different mass and redshift dependence than at lower redshifts.  First off, because $\eta$ at low masses is independent of $M_*$, this results in a flatter $Z(M_*)$ up to our chosen cutoff $M_*$.  Also, because mass accretion rates are generally higher at high redshifts, and this is not mitigated by high $\eta$ values, this means that the predicted MZR is actually higher at high redshifts \citep[see also the analysis in][]{Langan19, Torrey19}.

Returning to the UVLF, one effect of the higher MZR is lower UV flux output owing to stellar populations with correspondingly higher metallicity, especially in the range $0.1-1.0\ Z_\odot$. We have verified that the stellar metallicities in the simulated galaxies roughly trace the gas-phase metallicities, with $\sim0.2$~dex scatter. Changing the stellar metallicity from $0.1\ Z_\odot$ to $1.0\ Z_\odot$ at $z=6-8$ lowers the galaxy UV luminosity by $\sim0.4$~mag \citep[see Fig.~3 of][]{Tacc18}. However, the simulated no-dust UVLF surpasses the observations. The reduced UV flux is therefore not the main reason for the simulated $z=8$ UVLF being low.

We thus examine in detail the second effect of a higher MZR on the UVLF, which is an increase in the amount of dust extinction. \tmp{Since the $z=6-8$ MZR in \simba\ is higher than the observations at $z=2$, simply applying the $A_V$--metal column density scaling of the Milky Way (see Section~\ref{sec:methods}) would result in too much dust extinction. However, the steep DTM--Z scaling as shown in the bottom panel of Figure~\ref{fig:MZR} lowers the $A_V$ values of the faint galaxies, leading to descent agreement between the simulated $z=6$ UVLF and the observations. At $z=8$, the DTM of most galaxies are lower than at $z=6$, but} the even higher metallicities seem to be too extreme, resulting in too much dust extinction that lowers the UVLF.  This suggests that our assumption of $\eta$ at $z\gg 6$ in very small galaxies may need some modification.  Doing so may also help alleviate the discrepancy in the faint-end mass function of \simba-25.  We leave it for future work to tune this more carefully.  We note that while our assumed $\eta(M_*)$ is taken from that predicted in the FIRE simulations \citep{Angles17b}, FIRE does not offer strong guidance for this quantity at these very early epochs.

% \tmp{In addition to the MZR being high, the simulated DTM in \simba\ also appears to be higher than, e.g. \citet{popping17}. This also contributes to the excessive dust extinction at the bright end of the $z=8$ UVLF. Given the lack of observational constraints on the high-$z$ DTM, we leave a detailed exploration of the dust model to future work.}

Focusing on $z\sim 6$, \tmp{the fact that the high MZR predicted by \simba\ is able to match the observed UVLF and $\beta-\MUV$ relation indicates the possibility of early metal enrichment. Whether such early metal enrichment is necessary to reproduce the $z\sim6$ observations is slightly sensitive to the DTM ratio. If the sub-grid model locks significantly more metals into dust, the MZR would be lower and we may still be able to reproduce the UVLF. However, any significant reduction in the MZR would require very large DTM ratios that are not predicted by \simba~\citep{Li19}, nor expected owing to the time taken to grow dust grains from metals. Future \jwst\ observations will be able to constrain the metal enrichment history of these reionization-epoch galaxies with near-IR spectroscopy.}

\subsection{Galaxy Size and Morphology}
\label{sec:size}

A unique new capability of \jwst\ will be to able to resolve typical reionization-epoch galaxies in the rest-frame optical, as is evident from Figure~\ref{fig:example_images}.  With \hst, this is only possible in the rest-frame UV, and for fairly sizeable galaxies unless they are lensed.  It is thus interesting to make predictions for \jwst\ to understand the relationship between rest-UV and rest-optical sizes.  We do so here by making mock images with {\sc Pyloser} and mimicking the observational procedure to determine galaxy sizes.  Along the way we will also compare to available \hst\ observations of $z\approx6$ galaxy sizes.
 
We focus on sizes in the \jwst\ F115W and F444W bands, which correspond to rest-frame 1600 {\AA} (UV) and 6300 {\AA} (optical) at $z=6$, respectively\footnote{\tmp{We choose to focus on F115W simply because it is close to rest-frame 1500 {\AA} at $z=6$. We note that JWST/NIRCAM is under-sampled in F115W, so F150W or F200W may be a better choice for size measurements. However, the under-sampling problem can be mitigated by drizzling. Moreover, given the similarity of galaxy morphologies in F115W and F444W as presented in Section~\ref{sec:size}, we anticipate that using F150W or F220W instead of F115W will not affect our main conclusions.}}. The former is similar to the \hst\ F110W band, and we obtain very similar sizes in these two bands when extracted at the same resolution.  For each simulated galaxy, we generate mock \jwst\ images in these two bands with a side length of four times the maximum radius of the star particles to the galaxy's center of mass. Given that the \citet{Calz00} extinction law produces the overall best match of the simulations to the observed UVLF and $\beta-\MUV$ relation at $z=6$, we only use the \citet{Calz00} law to generate mock images. We perform size measurements using three methods listed below:

\begin{itemize}
\item {\it Growth curve method:} We generate both the F115W images and the F444W images using the F115W pixel scale, 0.031 arcsec. F444W images created in this way are denoted as high resolution (F444W-HR). \tmp{We compute the flux weighted center of the images and perform circular aperture photometry with the Python {\tt photutils}\footnote{\url{https://photutils.readthedocs.io}} package to get the growth curve. This is used to measure the half-light radius which contains half of the total light of the galaxy.}

\item {\it Sersic fit method:} We produce the F115W and F444W images using their original pixel scales, $0.031$ and $0.063$ arcsec, respectively.\footnote{We note that in real observations people usually combine a few tens of exposures, thereby drizzle them onto a smaller grid. Thus in reality the drizzled pixel scales can be smaller by a factor of about 2 than the original pixel scales ($0.031$ and $0.063$ arcsec) used in this work.} The images are then convolved with the PSFs of the two bands\footnote{\url{https://jwst-docs.stsci.edu/display/JTI/NIRCam+Point+Spread+Functions}}. We add Gaussian noise with a standard deviation that corresponds to $1/50$ of the median flux of the image. \tmp{This choice of the noise level is arbitrary, but does not affect our size measurement much because a larger effect comes from the PSF. We then use {\tt photutils} to produce a segmentation map, which is an integer-valued 2D array where pixels belonging to the same source are labelled with the same integer value. The background is tagged as zero in the segmentation map. We only keep the source with the largest area and smooth the main source segment by convolving the segmentation map with a uniform boxcar filter spanning 4 pixels in each dimension. This segmentation map and the PSF are fed into} the {\tt statmorph} package\footnote{\url{ https://statmorph.readthedocs.io}} to fit a Sersic profile \citep{Sersic68} to the galaxy image and obtain the half-light radius $r_e$ \citep{RG19}. The Sersic profile parametrizes the surface brightness profile of a galaxy as
\begin{equation}
\Sigma(r) = \Sigma_0 \exp\left[ -b_n \left( \frac{r}{r_e} \right)^{1/n} \right]
\end{equation}
where $\Sigma_0$ is the surface brightness at $r=r_e$, $b_n$ is a coefficient chosen so that a circular aperture with radius $r_e$ contains half of the galaxy's flux, and $n$ is the Sersic index. {\tt Statmorph} also takes a segmentation map and the PSF as input. During each step of the Sersic profile fitting, the modelled Sersic profile is convolved with the PSF to correct for the PSF effects. We keep all galaxies whose ``bad measurement" flags are 0, indicating good measurements \citep[{\tt flag} and {\tt flag\_sersic}; see][for details]{RG19}. This is a much looser selection criterion than that used in \citet{RG19} because we would like to keep more data. The ``good fit" criteria depend on other input parameters as well, such as the gain (which we set to 1). We therefore avoid removing a large fraction of galaxies out of our sample so as not to bias our results. However, we caution that $r_e$ values that fall below the PSF FWHM of the filters are unlikely to be reliable.

\item {\it 3D half-light method:} We calculate the fluxes in F115W and F444W for each star particle in a galaxy and compute a light-weighted centre. We then measure the three-dimensional half-light radius for the galaxy, which can be seen as the ``intrinsic'' size of the galaxy. \tmp{The dust extinction of each star particle is calculated along the same line of sight as the one adopted when producing the mock images. While in this case a 2D half-light radius seems better defined, in practice we find that the differences between the 2D and 3D sizes are only at $\sim20\%$ level. Therefore our analysis are not significantly affected by the choice of definition of the sizes.}
\end{itemize}

\begin{figure*}
	\includegraphics[width=2\columnwidth]{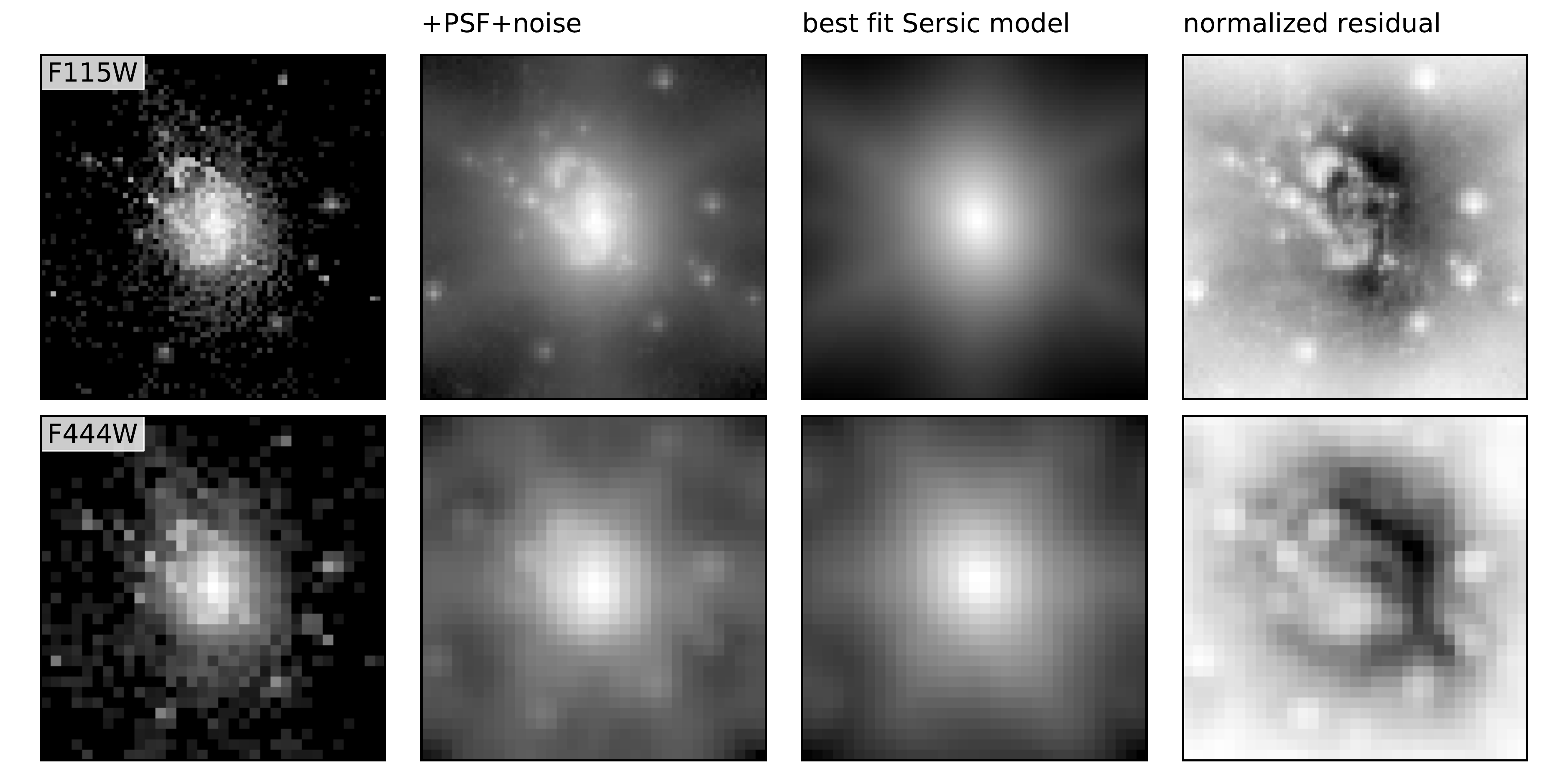}
    \caption{An illustration of our image generation and Sersic fit pipeline. Top and bottom panels show mock \jwst\ F115W and F444W images respectively. The leftmost column presents the original images of a galaxy in \simba-25. The images are centered at the center of mass of the galaxy, and have a side length of four times the 3D stellar radius of the galaxy. The second column shows images convolved with PSFs in the corresponding bands and added Gaussian noise with a standard deviation corresponding to $1/50$ of the median flux of the image. The third column illustrates the Sersic fit results, with the PSFs convolved. The normalized residual of the Sersic fit are shown in the rightmost column.}
    \label{fig:example_sersic}
\end{figure*}

Figure~\ref{fig:example_sersic} illustrates the steps of the Sersic fit method. Top and bottom panels show images in F115W and F444W respectively. The images are shown out to twice the 3D stellar radius of the galaxy, which is the largest distance of a star particle to the galaxy's center of mass as determined by {\sc Caesar}. The leftmost column shows the original images of the same galaxy in \simba-25 as shown in Figure~\ref{fig:example_images}. The second column presents the images after being convolved with PSFs in the two bands and added Gaussian noise. The best-fit Sersic profile models convolved with PSFs are shown in the third column. Finally, the rightmost column shows the normalized residual of the Sersic fit. The `bad measurement' flags of this galaxy all indicate successful Sersic fits, but not all the features can be fully fitted.  This is not surprising given that early small galaxies are often fairly irregular and clumpy in \simba, as is also the case in observations.  Nonetheless, the Sersic fit captures much of the overall profile.

\subsubsection{Size--luminosity relation}

\begin{figure*}
	\includegraphics[width=2\columnwidth]{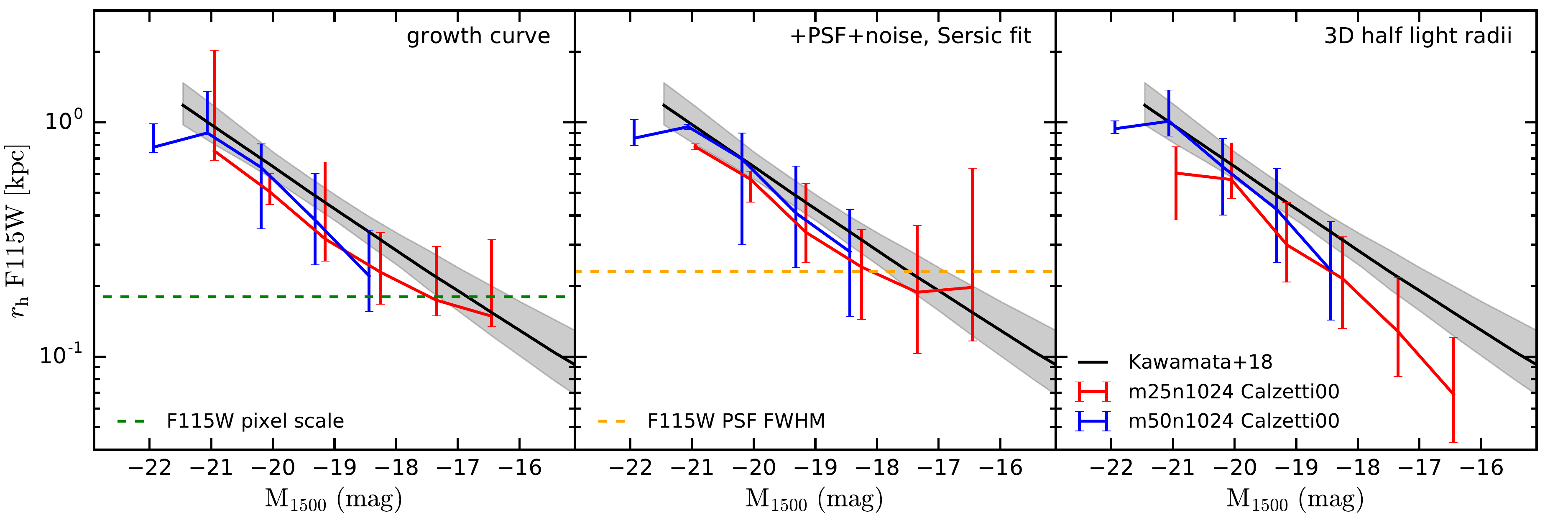}
    \caption{The size--UV luminosity relations measured in the \jwst\ F115W band (rest-frame UV) at $z=6$. From left to right, sizes are obtained using the growth curve method, the Sersic fit method, and the 3D half-light method respectively (see text for detaills). Red and blue represent results obtained from \simba-25 and \simba-50, respectively. Errorbars show $1\sigma$ spread in the sizes. For the growth curve and Sersic fit methods, we generate images assuming the \protect\citet{Calz00} law. For the 3D half-light method, the size measurements shown use the dust-attenuated flux in F115W. All methods yield a descent match to the observed size--luminosity relation.}
    \label{fig:size-luminosity}
\end{figure*}

Figure~\ref{fig:size-luminosity} shows the (rest-UV) size--luminosity relations at $z=6$ obtained using the growth curve method, the Sersic fit method, and the 3D half-light method from left to right. The sizes are measured in the F115W band and shown in units of physical kpc, and plotted versus $\MUV$. Red and blue lines illustrate results in \simba-25 and \simba-50, respectively. Errorbars represent the $1\sigma$ spread in the sizes. In the left and middle panels, the green and orange dashed lines show the F115W pixel scale and the PSF FWHM, respectively. The black lines show the observed size--luminosity relation in \citet{Kawa18}, and the shaded region represent $1\sigma$ error. \citet{Kawa18} measured the sizes of a sample of galaxies obtained in the {\it Hubble} Frontier Fields program by correcting the lensing effects and fitting Sersic profiles to the stacked images in the WFC3/IR bands.

The simulated size--luminosity relations are reasonably converged between the two simulations, using all three methods. Given that the nominal resolution of \simba-50 is about 50 physical pc at this redshift, and half that in \simba-25, and the F115W pixel scale is $\sim 4\times$ larger than this, it is not surprising that the predictions are robust to resolution effects, at least down the sizes probed by \jwst.

The galaxy sizes measured using the two observational approaches (growth curve and Sersic) agree well with \hst\ observations.  The brightest galaxies have sizes of about a physical kpc, whereas galaxies at $\MUV\approx -17$ to $-18$ approach the instrumental resolution; below this, it will be difficult to robustly measure sizes without the aid of lensing.  Sizes measured with growth curve method that fall below the F115W pixel scale (green dashed line) are not fully trustable. Similarly, sizes obtained with the Sersic method should be trusted when they are larger than the PSF FWHM (orange dashed line). \tmp{The 3D half-light radii agree with the two observational methods for the bright galaxies. The sizes of faint galaxies show a steeper decline than the observational data, indicating potential systematic errors in observations below the instrumental resolution limits.}

\begin{figure*}
	\includegraphics[width=2\columnwidth]{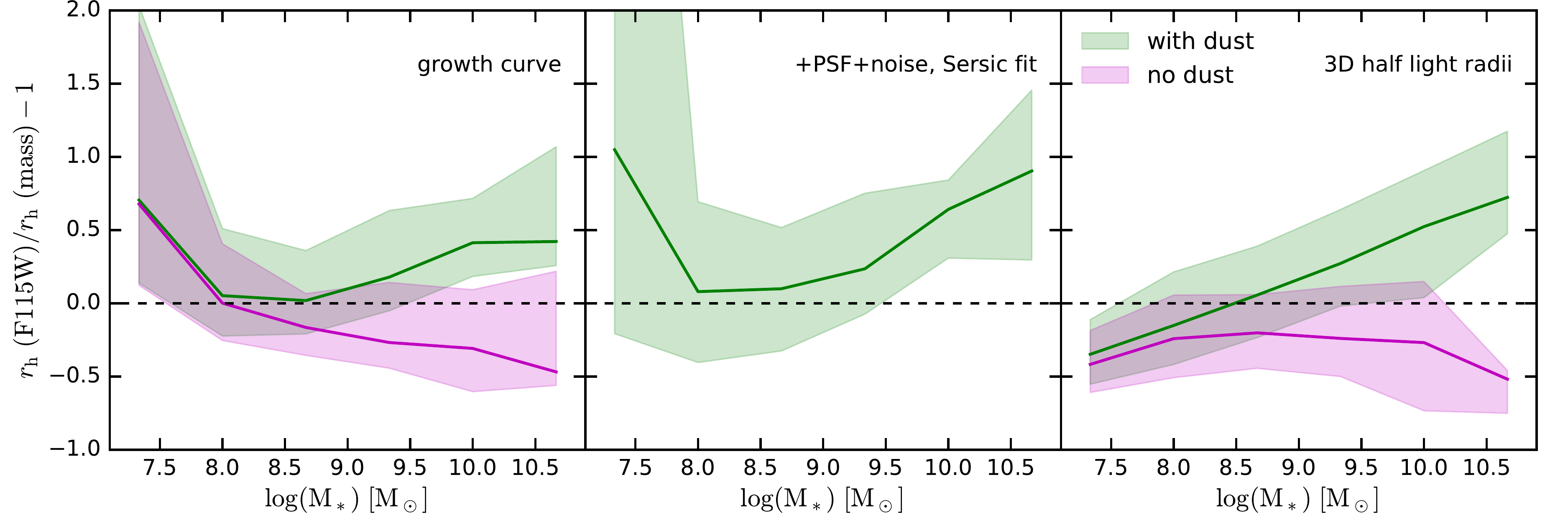}
    \caption{Similar to Fig.~\ref{fig:size-luminosity}, but showing comparisons of the measured F115W sizes (rest-UV) to the galaxy 3D half-stellar mass radii as a function of stellar mass. Shaded regions represent $1\sigma$ spread in the size ratios. The 3D half-light radii traces the half-mass radii very well at $z=6$. Due to the finite pixel size and smoothing by the PSF, size measurements in real observations bias the ``true'' sizes, which we find to be a mass-dependent effect.}
    \label{fig:delrh-M}
\end{figure*}

We examine how well the measured sizes using the three different methods recover the true 3D half-stellar mass radii. Figure~\ref{fig:delrh-M} illustrates the ratio of the F115W (rest-UV) sizes to the 3D half-stellar mass radii as a function of stellar mass, with the F115W sizes obtained by the three methods. \tmp{We have combined sampled from \simba-25 and \simba-50.} Solid green lines represent the median of the ratios, and the shaded regions show $1\sigma$ spread in the size ratios. \tmp{For the growth curve method and 3D half-light radii, we also show results using the no-dust sizes with magenta lines.}

\tmp{Focusing first on the right panel of Figure~\ref{fig:delrh-M}, the no-dust 3D half-light sizes are $20-50\%$ smaller than the 3D half-stellar mass radii, but the ratios are roughly constant. This indicates that although more concentrated than mass, light traces mass well in these simulated $z=6$ galaxies, so these galaxies can only have weak age gradients. When including dust extinction, the sizes could be a factor of $1.5-3$ larger than sizes without dust for galaxies with $M_*\gtrsim10^9\ M_\odot$, and also overestimate the half-mass radii by up to a factor of $\sim1.5$. This suggests that the centers of the galaxies are more dust-attenuated. Indeed, we will show in Section~\ref{sec:color} that the centers of these galaxies are younger but with more dust extinction. Results obtained with the two observational methods are generally similar, as shown in the left and middle panels. The sizes of galaxies with $M_*<10^8\ M_\odot$ are overestimated by factors of $1.5-2$, owing to the finite resolution of the mock images and smoothing of the PSF.}

\tmp{Overall, \simba\ produces galaxy sizes that are in good agreement with current \hst\ observations at $z=6$. We also find that the observed sizes are a biased tracer of the true half-stellar mass radii. Moving forward, we will explore predictions for \jwst\ observations.}

\subsubsection{Rest-optical vs rest-UV sizes}
\label{sec:f444w_size}

\begin{figure*}
	\includegraphics[width=2\columnwidth]{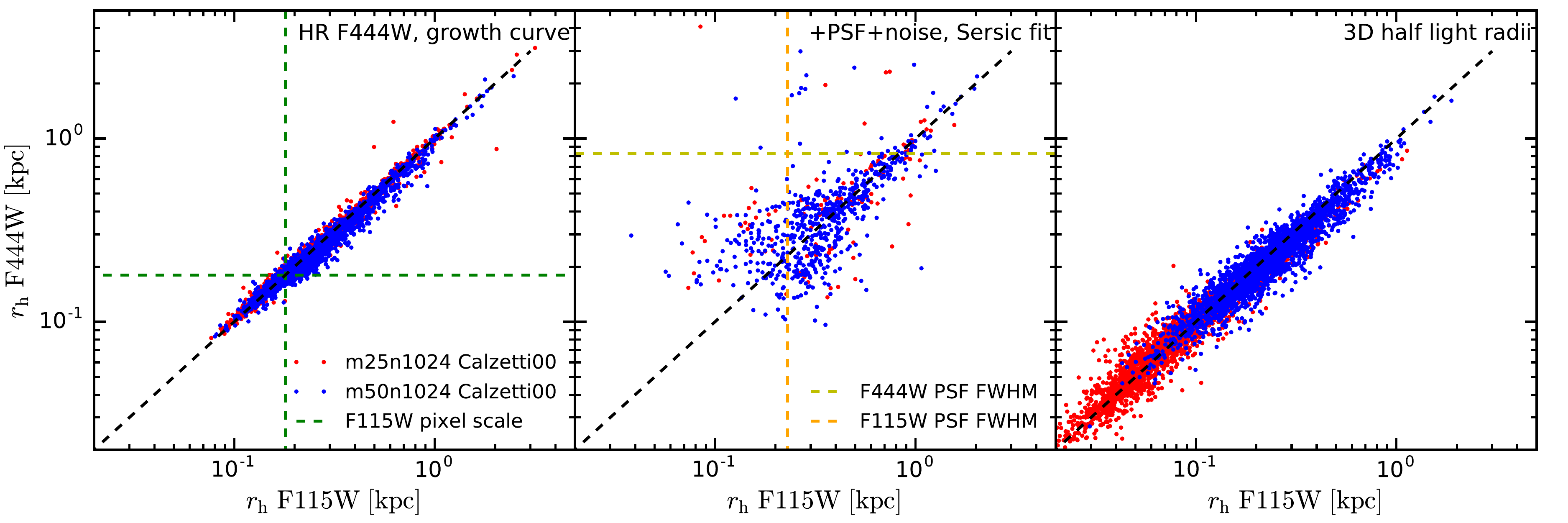}
    \caption{Similar to Fig.~\ref{fig:size-luminosity}, but showing comparisons between F444W (rest-optical) sizes and F115W (rest-UV) sizes at $z=6$. The green dashed lines in the left panel show the pixel scale of \jwst\ F115W. The orange and yellow dashed lines in the middle panel represent the FWHM of the PSFs in F115W and F444W, respectively. We did not remove galaxies with F444W sizes smaller than the PSF FWHM, but we caution that such size measurements may not be reliable. Regardless of the size measurement method used, the F115W (rest-UV) and F444W (rest-optical) sizes are very similar.}
    \label{fig:size_2bands}
\end{figure*}

Since future \jwst\ observations will be able to explore the $z\sim6$ galaxies in the rest-frame optical wavelengths, we explore how the rest-frame optical sizes compare with the rest-frame UV ones. Figure~\ref{fig:size_2bands} shows comparisons between the F444W (rest-optical) sizes and the F115W (rest-UV) sizes, measured using the growth curve method, the Sersic fit method, and 3D half-light method from left to right respectively. The green dashed line in the leftmost panel indicate the F115W pixel scale. The orange and yellow dashed lines in the middle panel represent the PSF FWHM of F115W and F444W respectively. In the rightmost panel showing the intrinsic 3D half-light radii, the F115W and F444W sizes match each other very well. This is surprising because for galaxies at much lower redshifts where the cores are redder and the disks are bluer, the rest-frame optical sizes should be smaller than the rest-frame UV sizes \citep[e.g.][]{Szom13, Lang14, Mosl17, Tacc15a, Tacc15b}. Our size measurements at $z=6$ therefore indicates young stellar populations and small age gradients with radius in these high-$z$ galaxies. \tmp{We note that although the sizes presented here include effects of dust extinction, the no-dust sizes also show a good one-to-one relation between rest-UV and rest-optical.}

The sizes measured with the two observational approaches make very similar predictions as the theoretical method on the sizes in rest-frame UV and optical. The growth curve method is limited by the finite pixel size, but still produces a clear one-to-one trend (shown by the black diagonal line) of the sizes in the two bands. \tmp{Our finding slightly disagrees with \citet{Ma18b}, who also measured sizes with mock images but found that the FIRE-2 $z=6$ galaxies have very clumpy UV morphologies and that the rest-optical traces the stellar mass better than UV. However, their galaxies are fainter ($\MUV>-18$) than ours. Moreover, the FIRE-2 galaxies have been shown to have bursty star formation histories owing to their star formation feedback model \citep{muratov15, sparre17, Ma18a}, making them more likely to have clumpy morphologies than the \simba\ galaxies.} The Sersic fit method gives more scatter in the measured sizes, primarily owing to the smoothing by the PSFs. The similarity in the F115W and F444W sizes can still be seen. However, because of the large FWHM of the F444W PSF ($0.83$ physical kpc at $z=6$) compared to the physical sizes of the galaxies (of order $\sim1$ kpc), a lot of the Sersic fit results go below the FWHM of the F444W PSF (horizontal dashed yellow line). The $z\sim6$ galaxies thus still require lensing in order for their size measurements in the rest-frame optical bands to be reliable with the observations of \jwst.

\subsubsection{Color, dust, and age gradients}
\label{sec:color}

\begin{figure*}
	\includegraphics[width=2\columnwidth]{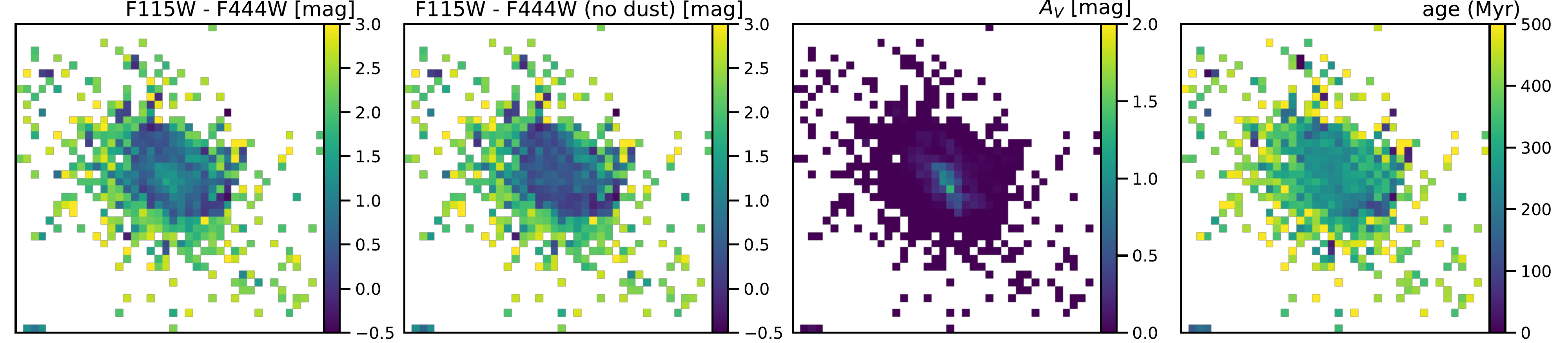}
	\includegraphics[width=2\columnwidth]{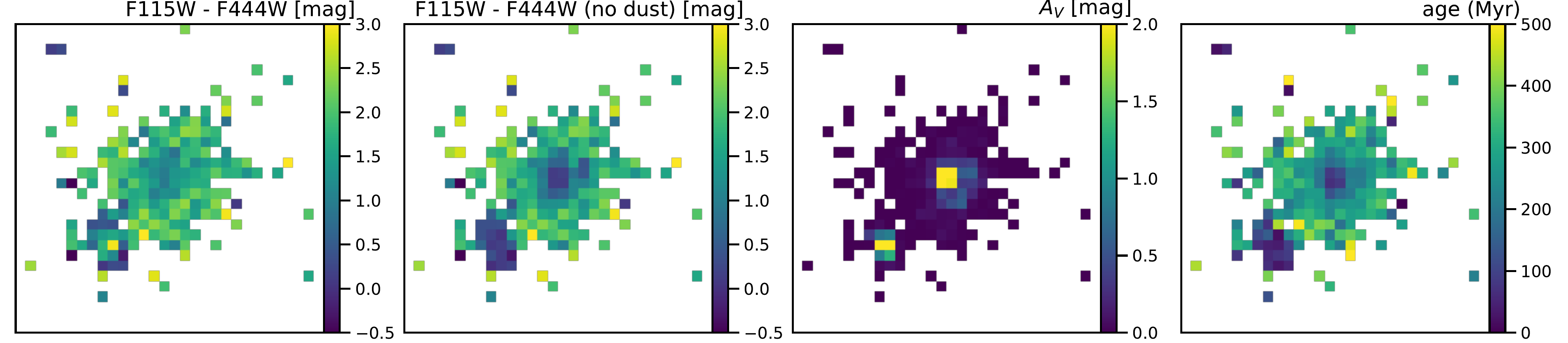}
    \caption{Examples of maps of \jwst\ F115W - F444W colors with (leftmost column) and without dust (second column), $A_V$ (third column), and age (rightmost column), created with two $z=6$ galaxies in \simba. These maps do not include PSF effects. The galaxy shown in the top row has $M_* = 2.3\times10^9\ M_\odot$ and SFR$=7.4\ M_\odot/{\rm yr}$, and the one in the bottom row has $M_* = 4.9\times10^9\ M_\odot$ and SFR$=40\ M_\odot/{\rm yr}$. The maps have the same pixel scale as F115W, $0.031$~arcsec ($0.18$~proper kpc at $z=6$). These maps indicate weak age gradient which makes the center of the galaxies bluer, while more dust in the center makes it redder and thus weakens the color gradient of the galaxies.}
    \label{fig:example_maps}
\end{figure*}

We now explore the age and $A_V$ gradients in the simulated galaxies in detail. Figure~\ref{fig:example_maps} presents maps of the \jwst\ F115W - F444W (rest-UV - rest-optical) color, $A_V$, and age, created for two galaxies in \simba. The maps have the same pixel size as F115W, $0.031$~arcsec. The galaxy shown in the top row has $M_* = 2.3\times10^9\ M_\odot$ and SFR$=7.4\ M_\odot/{\rm yr}$, and the one in the bottom row has $M_* = 4.9\times10^9\ M_\odot$ and SFR$=40\ M_\odot/{\rm yr}$. The maps are shown out to twice the 3D stellar radius of the galaxy, which is the largest distance of a star particle to the galaxy's center of mass as determined by {\sc Caesar}. 

With dust extinction, the galaxies exhibit roughly uniform F115W - F444W color (leftmost column), with the top galaxy having a slightly redder core and the bottom one having a slightly bluer core.  Without dust, the core is clearly bluer in the center (second column), especially for the bottom galaxy.  Both galaxies have $A_V$ maps that peak in the center (third column), as expected since this is where the densest gas resides, and also the metallicity is higher in the core.  The effect is stronger for the bottom galaxy.  Also stronger for the bottom galaxy is the age gradient (rightmost column), with a very young core compared to the top one, which is perhaps related to the fact that the bottom galaxy has a three times higher sSFR.  Interestingly, the galaxy in the bottom row shows an off-centre star-forming clump near the lower-left corner which also has high $A_V$ values, suggesting that giant clumps as seen in $z\sim 2$ galaxies may also be evident at these epochs.  Curiously, the combined impact of the younger core with higher $A_V$ seems to approximately cancel out in the F115W - F444W color, leaving a mildly bluer core in the bottom galaxy.  In the top galaxy, there is essentially no young core, and a weak $A_V$ gradient, leading to a slightly redder core. 

These results indicate that the weak age gradients of the simulated $z=6$ galaxies make the centers of the galaxies bluer, while more dust in the central star-forming regions makes them redder. The two effects weaken the color gradients of the galaxies, leading to very similar rest-UV and rest-optical sizes. \tmp{This finding also explains the differences among the sizes with dust, without dust, and the stellar half-mass radii.} Breaking this well-known age-dust degeneracy would ideally require longer wavelength data with \jwst/MIRI, although such data will be more limited in sensitivity and resolution as compared with NIRCAM data.  These maps demonstrate how reionization-epoch galaxies may be different from those at lower redshifts, which might be testable with future \jwst\ observations.

\tmp{Finally, we briefly comment on how the adopted DTM affects the sizes and morphologies of the simulated galaxies. Since the DTM increases with metallicity in \simba\ (Figure~\ref{fig:MZR}), the higher metallicity regions, i.e. the star-forming central parts of the galaxies, are more dust-attenuated than what would be expected assuming constant DTM. Therefore if we measured the sizes using constant DTM, the differences between the sizes with and without dust would be lower (Figure~\ref{fig:delrh-M}). However, given the age gradients in the simulated galaxies are small, we expect that the rest-UV and rest-optical sizes would still be very similar if assuming constant DTM (Figure~\ref{fig:size_2bands}). We note that we have implicitly assumed that dust tracks metals, and that in principle one would prefer using dust radiative transfer to fully take into account of the dust distribution. Moreover, several authors \citep[e.g.][]{kli16, Ma19} have pointed out the importance of correctly modeling dust scattering when comparing simulations to observations. These aspects will be explored in detail in future work.}

In summary, \simba\ produces very similar rest-optical and rest-UV sizes at $z=6$, and thus predicts that early galaxies do not exhibit the more concentrated mass distribution relative to UV light as observed at lower redshifts. Future \jwst\ observations may be able to test our prediction of the central dust-obscured star formation and weak age gradients in the $z=6$ galaxies, which differ from the findings of galaxies at lower redshifts.

\section{Summary}
\label{sec:conclusions}

In this work, we study the photometric properties of reionization-epoch galaxies in the cosmological hydrodynamic simulations from the \simba\ suite having $25\hmpc$ (\simba-25) and $50\hmpc$ (\simba-25) volumes. \simba-25 reaches resolutions $\la 50$~pc physical at $z=6$ and resolves all hydrogen cooling halos, and \simba-50 exhibits good resolution convergence in all properties while allowing larger objects to be represented.  Our main results are highlighted below:

\begin{itemize}
    \item \simba\ produces reasonable agreement with the observed $z=6$ UV luminosity function, owing to a partial cancellation between the relatively high metallicities in the simulated galaxies and a steep dust-to-metal ratio--gas-phase metallicity relation. We find that the UVLF at $\MUV\gtrsim-21$ is not sensitive to the form of the dust extinction law assumed. \simba\ under-produces the UVLF at $z=8$, which may be caused by too few massive galaxies but more likely because of too much metal that increases the amount of dust extinction.
    
    \item When assuming a \citet{Calz00} extinction law, \simba\ produces a relationship between the UV continuum slope $\beta$ and $\MUV$ in very good agreement with available observations at $z=6$, with a hint that the shallower \citet{Salm16} modification is preferred for faint galaxies. We find that $\beta$ is a good tracer of the dust extinction law, with the \citet{Salm16} law and the SMC law producing $\Delta\beta\sim0.8$. Without dust, the simulated $\beta-\MUV$ relation is flat, owing to the simulated star formation rate--stellar mass relations at $z=6-8$ having a slope of $\approx1$. This results in similar amounts of time to double the stellar masses in different galaxies, thus similar ages and metallicities of the stellar populations.
    
    \item The gas-phase metallicities of \simba\ galaxies at a given $M_*$ are higher at $z=6$ than at $z=2$. At $z=8$, \simba\ predicts an even higher mass-metallicity relation. The higher metallicity affects both the UV flux output of stellar populations and the amount of dust attenuation, but more significantly the latter. This metallicity evolution is understandable within a simple scenario where the gas-phase metallicity reflects a balance of pristine inflow versus enriched outflow, given the assumed mass outflow rates in \simba. A more fine-tuning of the mass loading factor in low-mass galaxies may mitigate the tension between the low-mass end of simulated stellar mass function and the observations at $z=6$, as well as the simulated UVLF and the observations at $z=8$. Nevertheless, the fact that the simulated $z=6$ UVLF and $\beta-\MUV$ relation match the observations hints upon the possibility of early enrichment in high-redshift galaxies, which will be testable with future \jwst\ observations.
    
    \item We measured the sizes of the simulated $z=6$ galaxies by generating mock \jwst\ images in the F115W (rest-UV) and F444W (rest-optical) bands. We find that \simba\ produces F115W size--$\MUV$ relations that are in reasonable agreement with the observations of \hst, regardless of whether the size measurement mimics the methods used in real observations, or is performed by theoretical calculations.% We find that galaxies with $\MUV\gtrsim-18$ have half-light radii that fall below the F115W pixel scale, or the FWHM of the PSF. These galaxies thus still require lensing for a robust measurement of their sizes in the \jwst\ era.
    
    \item We find that the simulated galaxies have very similar sizes in rest-UV and rest-optical, regardless of the method of size measurement. This has great implications for \jwst\ because it will be able to observe the $z=6$ galaxies in the rest-optical bands. Most simulated galaxies have sizes below the FWHM of the F444W PSF, indicating that lensing is needed for their size measurements to be trustable.
    
    \item \tmp{We find that without dust extinction, light traces mass well in the simulated galaxies, owing to weak age gradients and young stellar populations. With dust extinction, the galaxy sizes can be factors of $1.5-3$ larger than the stellar half-mass radii for $M_*\gtrsim10^9\ M_\odot$, since the highly star-forming regions are also the most dust-extincted. The observed sizes can thus be a biased tracer of the half-mass radii.}
    
    \item We find that the simulated galaxies show younger ages and more dust-obscured star formation in the inner parts, with age gradients accompanying similar extinction gradients. These two effects approximately counteract each other to produce weak F115W-F444W color gradients in the simulated $z=6$ galaxies, leading to very similar rest-UV and rest-optical sizes. This prediction is very different from the findings of the color gradients of galaxies at lower redshifts, and thus will be interesting to test with future \jwst\ observations.
\end{itemize}

Future observational facilities including \jwst\ will be able to yield more accurate measurements of $z\gtrsim6$ galaxies, such as the stellar population properties and the enrichment histories. Accompanying multi-wavelength observations such as with ALMA will be crucial for providing an independent measure into the gas content and star formation rate, thereby giving independent avenues for constraining physical properties if such measures can be robustly interpreted within galaxy formation models (e.g. Leung et al., in prep.).  Galaxy formation simulations such as \simba\ thus can provide promising directions for future observations to probe, which in turn will act as direct tests of the input physical models. It is of particular interest to compare the predictions from simulations with different galaxy formation models. For instance, the Illustris-TNG50 simulation has a comparable volume and resolution as \simba, and allows for a statistical examination of the galaxy population properties \citep{Nelson19, Pill19}. Such simulations can also provide a rich database for a thorough mimicking of the observational methods in order to understand the bias between the observed quantities and the underlying physical properties. \tmp{It would also be interesting to perform full dust radiative transfer to test the robustness of the predictions presented in this work, which would make full use of the dust model in \simba\ and take into account important physical effects not considered here, such as dust scattering.}

\section*{Acknowledgements}

We thank Rachael Somerville, Aaron Yung, Daisy Leung, and Daniel Ang\'es-Alc\'azar for useful conversations regarding this work. This work was initiated as a project for the Kavli Summer Program in Astrophysics held at the Center for Computational Astrophysics of the Flatiron Institute in 2018. The program was co-funded by the Kavli Foundation and the Simons Foundation. We thank them for their generous support. S.T. is supported by the Smithsonian Astrophysical Observatory through the CfA Fellowship.  RD acknowledges support from the Wolfson Research Merit Award program of the U.K. Royal Society. The computing equipment to run \simba\ was funded by BEIS capital funding via STFC capital grants ST/P002293/1, ST/R002371/1 and ST/S002502/1, Durham University and STFC operations grant ST/R000832/1. DiRAC is part of the National e-Infrastructure.

%%%%%%%%%%%%%%%%%%%%%%%%%%%%%%%%%%%%%%%%%%%%%%%%%%



%\bibliographystyle{mnras}


%%%%%%%%%%%%%%%%%%%%%%%%%%%%%%%%%%%%%%%%%%%%%%%%%%

%%%%%%%%%%%%%%%%% APPENDICES %%%%%%%%%%%%%%%%%%%%%

\appendix

\section{Comparing different isochrone models}
\label{sec:compare_bpass}

We examine how the simulated UVLF and $\beta-\MUV$ relation change with different isochrone models. Specifically, we run {\sc Pyloser} with three different combinations of isochrone model and spectral library implemented in FSPS: MIST + MILES, PADOVA + MILES, BPASS (with pre-computed single stellar populations) \citep{bpass17, bpass18}. Because the BPASS model assumes the \citet{Salp55} IMF, we run the other models with this IMF for better comparison. Nebular emission (pre-computed in FSPS) is also included. We perform this calculation only for the $z=6$ galaxies in \simba-25

\begin{figure}
	\includegraphics[width=\columnwidth]{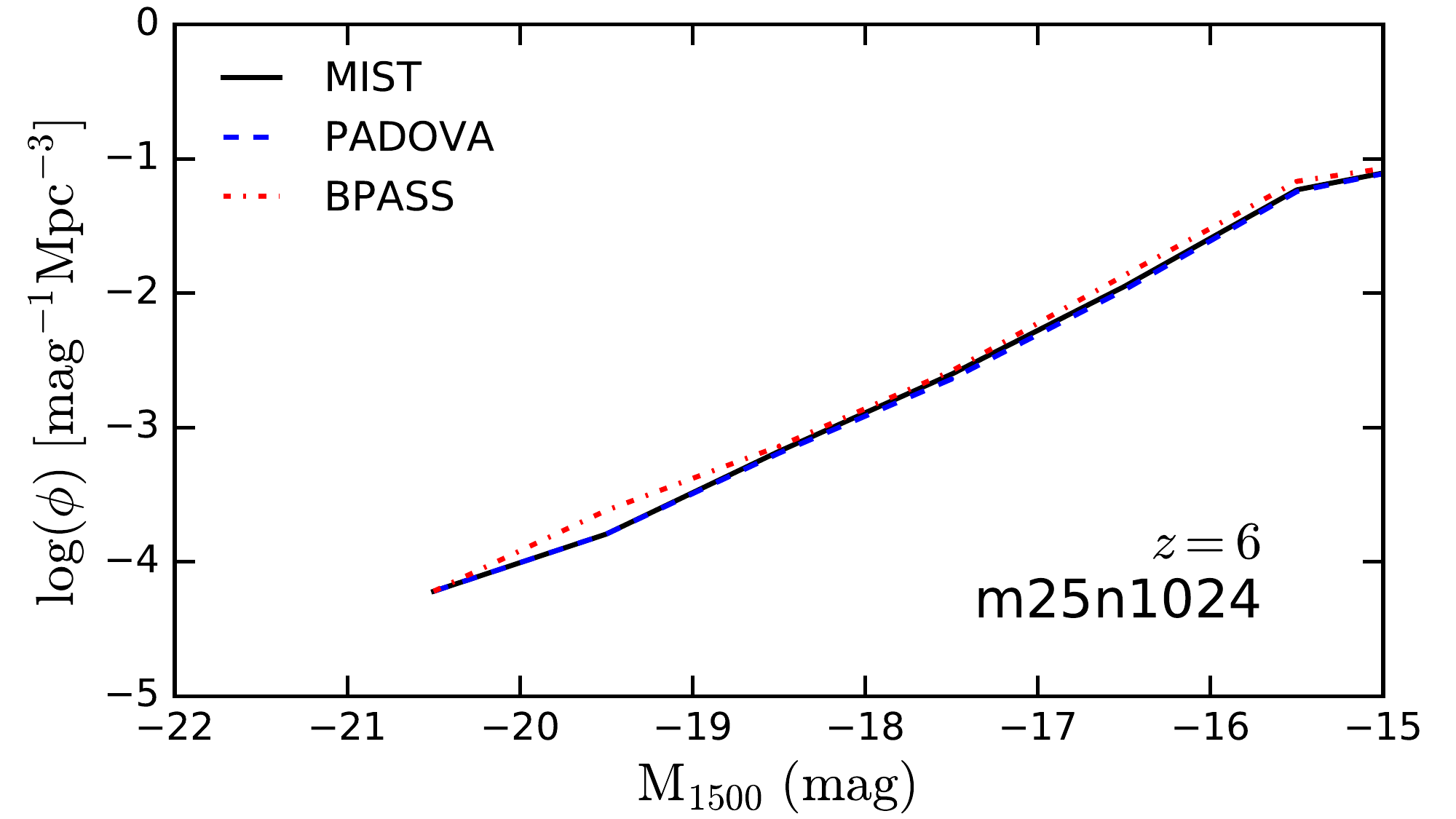}
	\includegraphics[width=\columnwidth]{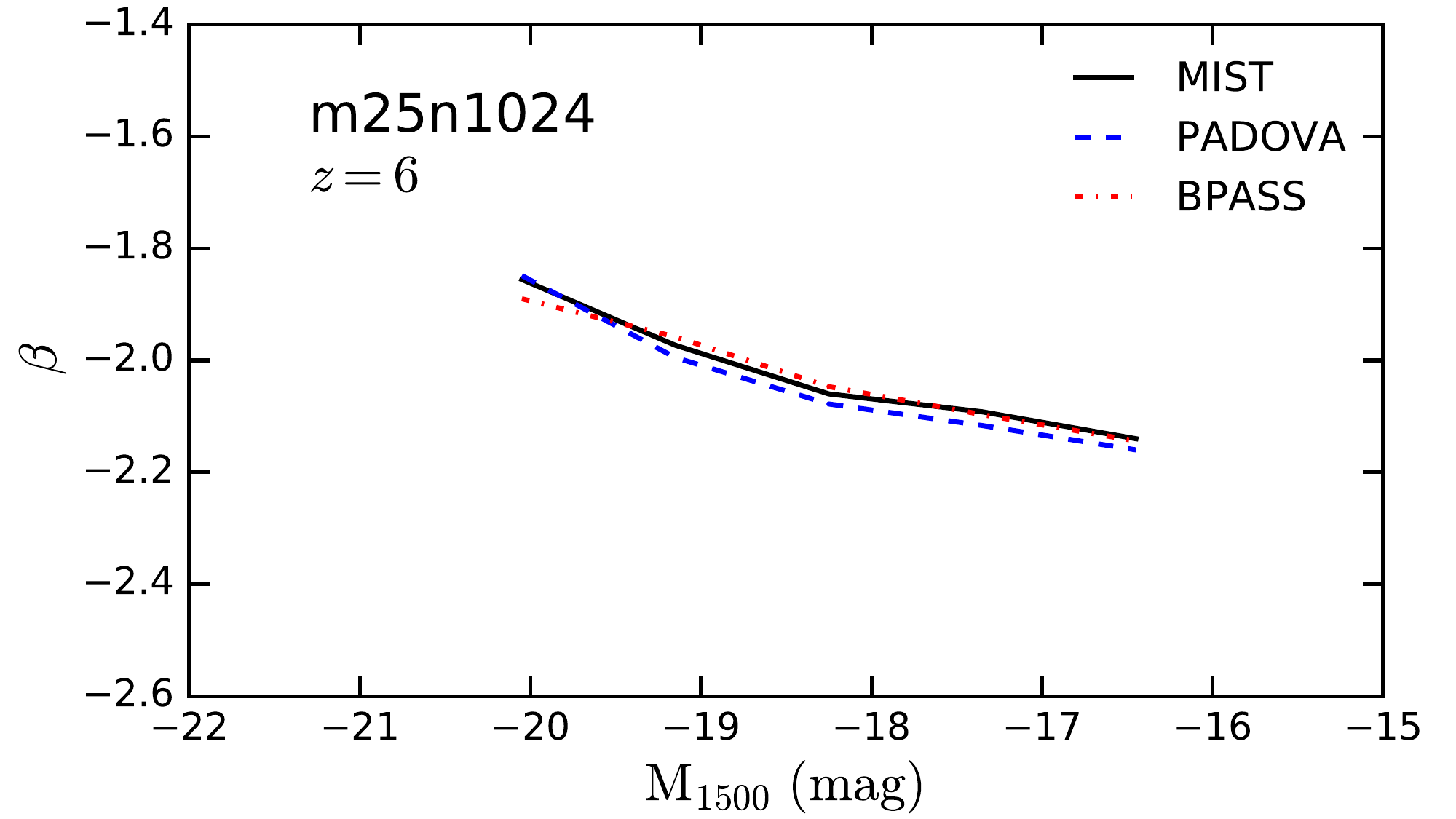}
    \caption{The UVLF (top) and $\beta-\MUV$ relation (bottom) at $z=6$ in \simba-25, computed using MIST + MILES (black solid lines), PADOVA + MILES (blue dashed lines), and BPASS (red dot-dashed lines). The predictions on the far-UV spectrum are very similar among different models.}
    \label{fig:compare_bpass}
\end{figure}

Figure~\ref{fig:compare_bpass} shows the UVLF (top panel) and $\beta-\MUV$ relation (bottom panel) at $z=6$ in \simba-25. Black solid, blue dashed, and red dot-dashed lines represent MIST, PADOVA, and BPASS respectively. The different model predictions on the UVLF and the $\beta-\MUV$ relation are thus very similar. This is consistent with the findings of \citet{Choi17} that the predictions on the far-UV spectrum by different isochrone models are very similar. It is only the ionizing photon production that differs a lot when changing the isochrone models. Thus the photometric properties of the simulated galaxies are robust with respect to changes in the isochrone model. This comparison also implies that altering an isochrone model cannot mitigate the disagreement between the simulated $z=8$ UVLF and the observations.

%%%%%%%%%%%%%%%%%%%%%%%%%%%%%%%%%%%%%%%%%%%%%%%%%%

% Don't change these lines
\bsp	% typesetting comment
\label{lastpage}
\end{document}